\documentclass[10pt,journal,compsoc]{IEEEtran}

\ifCLASSOPTIONcompsoc
  \usepackage[nocompress]{cite}
\else
  \usepackage{cite}
\fi

\ifCLASSINFOpdf
  \usepackage[pdftex]{graphicx}
  \graphicspath{{../pdf/}{../jpeg/}}
  \DeclareGraphicsExtensions{.pdf,.jpeg,.png}
\else
  \graphicspath{{../eps/}}
  \DeclareGraphicsExtensions{.eps}
\fi
\usepackage{subfigure}

\usepackage{amsmath,amssymb,amsfonts}
\usepackage{amsthm} 
\usepackage{algorithmic}

\ifCLASSOPTIONcompsoc
 \usepackage[caption=false,font=normalsize,labelfont=sf,textfont=sf]{subfig}
\else
 \usepackage[caption=false,font=footnotesize]{subfig}
\fi

\usepackage{threeparttable}

\usepackage{color}
\usepackage{bm}
\usepackage{float}
\usepackage{makecell} 
\usepackage{multirow}
\usepackage{booktabs} 
\usepackage{bigstrut}
\renewcommand{\thetable}{\arabic{table}} 

\usepackage{algorithm}
\usepackage{algorithmic}

\usepackage{amssymb} 

\newcommand{\tabincell}[2]{\begin{tabular}{@{}#1@{}}#2\end{tabular}}

\usepackage{mathtools}
\usepackage{letltxmacro}
\LetLtxMacro\orgvdots\vdots
\LetLtxMacro\orgddots\ddots
\makeatletter
\DeclareRobustCommand\vdots{%
  \mathpalette\@vdots{}%
}
\newcommand*{\@vdots}[2]{%
  \sbox0{$#1\cdotp\cdotp\cdotp\m@th$}%
  \sbox2{$#1.\m@th$}%
  \vbox{%
    \dimen@=\wd0 %
    \advance\dimen@ -3\ht2 %
    \kern.5\dimen@
    \dimen@=\wd2 %
    \advance\dimen@ -\ht2 %
    \dimen2=\wd0 %
    \advance\dimen2 -\dimen@
    \vbox to \dimen2{%
      \offinterlineskip
      \copy2 \vfill\copy2 \vfill\copy2 %
    }%
  }%
}
\DeclareRobustCommand\ddots{%
  \mathinner{%
    \mathpalette\@ddots{}%
    \mkern\thinmuskip
  }%
}
\newcommand*{\@ddots}[2]{%
  \sbox0{$#1\cdotp\cdotp\cdotp\m@th$}%
  \sbox2{$#1.\m@th$}%
  \vbox{%
    \dimen@=\wd0 %
    \advance\dimen@ -3\ht2 %
    \kern.5\dimen@
    \dimen@=\wd2 %
    \advance\dimen@ -\ht2 %
    \dimen2=\wd0 %
    \advance\dimen2 -\dimen@
    \vbox to \dimen2{%
      \offinterlineskip
      \hbox{$#1\mathpunct{.}\m@th$}%
      \vfill
      \hbox{$#1\mathpunct{\kern\wd2}\mathpunct{.}\m@th$}%
      \vfill
      \hbox{$#1\mathpunct{\kern\wd2}\mathpunct{\kern\wd2}\mathpunct{.}\m@th$}%
    }%
  }%
}
\makeatother

\makeatletter
\newcommand{\mathleft}{\@fleqntrue\@mathmargin0pt}
\newcommand{\mathcenter}{\@fleqnfalse}
\makeatother

\begin{document}

\title{An Efficient Matrix Multiplication with Enhanced Privacy Protection in Cloud Computing and Its Applications}

\author{Chun~Liu,
        Xuexian~Hu,
        Xiaofeng~Chen,~\IEEEmembership{Senior~Member,~IEEE,}
        Jianghong~Wei,
        Wenfen~Liu
\IEEEcompsocitemizethanks{\IEEEcompsocthanksitem C. Liu is with the State Key Laboratory of Mathematical Engineering and Advanced Computing, PLA Strategic Support Force Information Engineering University, Zhengzhou, 450001, China.
(e-mail: 3130100551@zju.edu.cn)
\IEEEcompsocthanksitem X. Hu (corresponding author) is with the State Key Laboratory of Mathematical Engineering and Advanced Computing, PLA Strategic Support Force Information Engineering University, Zhengzhou, 450001, China.
(e-mail: xuexian\_hu@hotmail.com)
\IEEEcompsocthanksitem X. Chen is with the State Key Laboratory of Integrated Service Networks (ISN), Xidian University, China and the State Key Laboratory of Cryptology, P.O. Box 5159, Beijing, 100878, China.
(e-mail: xfchen@xidian.edu.cn)
\IEEEcompsocthanksitem J. Wei is with the State Key Laboratory of Integrated Service Networks (ISN), Xidian University, China and State Key Laboratory of Mathematical Engineering and Advanced Computing, PLA Strategic Support Force Information Engineering University, Zhengzhou, 450001, China.
(e-mail: jianghong.wei.xxgc@gmail.com)
\IEEEcompsocthanksitem W. Liu is with the Guangxi Key Laboratory of Cryptogpraphy and Information Security, School of Computer Science and Information Security, Guilin University of Electronic Technology, Guilin, 541004, China.
(e-mail: liuwenfen@guet.edu.cn)}
}

\markboth{Journal of \LaTeX\ Class Files,~Vol.~14, No.~8, August~2015}%
{Liu \MakeLowercase{\textit{et al.}}: An Efficient Matrix Multiplication with Enhanced Privacy Protection in Cloud Computing and Its Applications}

\IEEEtitleabstractindextext{%
\begin{abstract}
As one of the most important basic operations, matrix multiplication computation (MMC) has varieties of applications in the scientific and engineering community such as linear regression, k-nearest neighbor classification and biometric identification. However handling these tasks with large-scale datasets will lead to huge computation beyond resource-constrained client's computation power. With the rapid development of cloud computing, outsourcing intensive tasks to cloud server has become a promising method. While the cloud server is generally out of the control of clients, there are still many challenges concerned with the privacy security of clients' sensitive data. Motivated by this, Lei et al. presented an efficient encryption scheme based on random permutation to protect the privacy of client's data in outsourcing MMC task. Nevertheless, there exists inherent security flaws in their scheme, revealing the statistic information of zero elements in the original data thus not satisfying the computational indistinguishability (IND-ZEA). 
Aiming to enhance the security of the outsourcing MMC task, we propose a new encryption scheme based on subtly designed invertible matrix where the additive perturbation is introduced besides the multiplicative perturbation. 
Furthermore, we show that the proposed encryption scheme can be applied to not only MMC task but also other kinds of outsourced tasks such as linear regression and principal component analysis. 
Theoretical analyses and experiments indicate that our methods are more secure in terms of data privacy, with comparable performance to the state-of-the-art scheme based on matrix transformation.
\end{abstract}

}

\maketitle

\IEEEdisplaynontitleabstractindextext

\IEEEpeerreviewmaketitle

\IEEEraisesectionheading{\section{Introduction}\label{sec:1}}

\IEEEPARstart{I}{n} the scientific and engineering community, \textit{matrix multiplication computation} (MMC) is one of the most important basic operations \cite{lei2014achieving} and has varieties of applications such as linear regression \cite{zhou2018efficiently}, k-nearest neighbor classification \cite{wong2009secure} and biometric identification \cite{zhou2018passbio}. However handling these tasks with large-scale datasets will lead to huge computation beyond small party's computation power, \textit{i.e.} resource-constrained client. 
For example, given a $m \times n$ matrix ${\bf{X}}$ and a $n \times s$ matrix ${\bf{Y}}$, the computation complexity of matrix multiplication ${\bf{Z}} = {\bf{X}}{\bf{Y}}$ is $O(mns)$.
With the rapid development of cloud computing, outsourcing intensive tasks to the cloud server has become an increasingly popular method for resource-constrained clients. They can leverage massive resources of cloud servers to complete these intensive tasks, typically based on a flexible pay-per-use manner \cite{wang2011secure,xu2014proof}.

Despite the positive prospect and trend in outsourcing intensive tasks to the cloud server, there are still several challenges concerned with privacy security, efficiency and verification. 
First, the input dataset and output result of the client may contain sensitive information (\textit{e.g.} traffic data or genomic data). Disclosure of this information to malicious attackers will endanger the safety of society and individual. Therefore, clients have to encrypt their data before outsourcing so that cloud servers could learn no additional information of the original dataset and the result. Though existing traditional encryption schemes can protect the sensitive data, these schemes affect the functionality of computation over the outsourced encrypted data. The promising homomorphic encryption makes it possible to solve this problem \cite{jiang2018secure}, but such schemes usually lead to a low computational efficiency far from practical \cite{gentry2009fully,shan2018practical}. So, encryption schemes based on matrix transformation are studied \cite{lei2014achieving,zhou2018efficiently,wong2009secure,zhou2018passbio}. 
Secondly, since the cloud server is not fully trusted by the client, there also needs to be a relatively robust method to verify the output result so that any false result can be resisted. 
Furthermore, besides the mentioned challenges, there is a strong need for a generic encryption scheme, which means it has a wide applicability and can be applied to not only MMC task, but also other kinds of outsourced tasks.

\subsection{Related work}

For privacy-preserving outsourcing schemes based on matrix transformation, random invertible matrices are generally used as the secret key to encrypt the original data \cite{shan2018practical}. 
While computing the inverse of an ordinary invertible matrix suffers heavy computation burden, \textit{e.g.} the computation complexity of computing the inverse of a $n \times n$ matrix ${\bf{O}}$ is $O(n^3)$.  
Using it to encrypt the original data (\textit{e.g.} ${\bf{X}}'={\bf{O}} {\bf{X}}$) or decrypt the result ciphertext (\textit{e.g.} ${\bf{Z}}={\bf{O}}^{-1} {\bf{Z}}'$) is also a heavy computation process for the client, which is actually a MMC process. 
Therefore, if the invertible matrix is not constructed subtly,
the computation complexity of the resource-constrained client to encrypt the sensitive data and decrypt the result ciphertext is comparable to that of cloud server, and the outsourcing doesn't make sense in consideration of significantly reducing the client's computation overload. 
Aiming for reducing the computation burden on the client side, Lei et al. designed a matrix encryption scheme based on random permutation and apply it to outsourcing large-scale MMC task \cite{lei2014achieving}. The main idea is to encrypt or decrypt the original large-scale matrix by multiplying special sparse matrices (\textit{i.e.} random permutation matrices). This encryption scheme can also be applied to matrix inversion computation (MIC) \cite{lei2013outsourcing} and matrix determinant computation (MDC) \cite{lei2014cloud}. Though the encryption scheme based on random permutation is efficient, it suffers from some privacy security weakness pointed out by \cite{fu2017secure,kumar2019secure,wang2018more,zhao2018sparse,zhang2019practical}, leaking the statistic information of zero elements in the original data. 
From the encryption scheme design perspective, this weakness results from only using multiplication perturbation without addition.
Motivated by this observation, Fu et al. proposed an improved protocol to outsource MMC based on matrix addition operation \cite{fu2017secure}. The random matrix is generated by multiplying a small-scale matrix with its transpose and then added to the original matrix. Nevertheless their encryption scheme can not be generalized to outsource other tasks such as MIC and MDC, let alone machine learning algorithms.
Based on Fu et al.'s work, Kumar et al. used the random permutation matrix to perturb the encrypted data further \cite{kumar2019secure}, but this method could also only solve the outsourced MMC task.
Employing more matrix addition and scalar multiplication operations, Wang et al. put forward a more secure outsource protocol for MMC \cite{wang2018more}, whereas it also can't be applied to other kinds of outsourced tasks and introduces more communication overhead and extra computation cost over the cloud server.
Except the direct matrix addition transformation, 
some work blended addition perturbation into the transformation based on matrix multiplication.
Zhao et al. proposed a sparse invertible matrix encryption scheme for outsourced computation, in which there are at most three nonzero elements in a row (\textit{resp.} a column) of the sparse matrix \cite{zhao2018sparse}. However, their encryption scheme can only solve the outsourcing task such as system of linear equations (LES) and MDC, where the client doesn't need to compute the inverse of the sparse invertible matrix to encrypt the data or decrypt the result. So, their scheme isn't suitable for outsourced MMC task.
Later, Zhang et al. gave out an outsource scheme of matrix operations including MMC, MIC and MDC based on unimodular matrix \cite{zhang2019practical}, where the data is perturbed first by random permutation and then by consecutive unimodular matrix transformations further. While the time cost of secret key generation would be increased. For example, to encrypt a $m \times n$ matrix ${\bf{X}}$, the client has to generate $m+n-2$ unimodular matrices and use them to perturb the sensitive data.

To discuss the applicability of matrix-transformation-based encryption scheme, two widely used machine learning algorithms \textit{linear regression} (LR) and \textit{principal component analysis} (PCA) are considered in this paper. Privacy-preserving outsourcing schemes based on matrix transformation of these two algorithms are as follows:

\subsubsection{linear regression}
Chen et al. proposed a perturbation approach by multiplying the original data with diagonal matrices \cite{chen2014highly}. However as mentioned in \cite{zhou2018efficiently}, there exists some limitations that Chen et al.'s scheme may be damaged by greatest common divisor attack and average absolute value attack. Moreover the number of zero elements is also leaked. 
In order to solve these limitations, Zhou et al. presented an encryption scheme to securely outsource LR to cloud server \cite{zhou2018efficiently}. They further gave out some other applications to show the applicability of their encryption scheme. Neverthless, it may be not suitable for the situation where encryption/decryption needs the inverse matrix of the secret key for the reason that computing the inverse matrix consumes huge computation resources comparable to those of cloud server.

\subsubsection{principal component analysis}
Based on random permutation, Zhou et al. constructed sparse orthogonal matrices to encrypt the original data of outsourcing PCA task \cite{zhou2016outsourcing}. Similar to Lei et al.'s encryption scheme, some inherent weakness would result in disclosure of sensitive data. Later using random orthogonal matrices, Liu et al. proposed a PCA outsourcing framework \cite{liu2018privacy}. However, generating a random orthogonal matrix as secret key and multiplying it with the original data themselves are heavy computation tasks for the resource-constrained client.

\subsection{Our contribution}
To further exploit the high performance of matrix transformation and remedy the inherent privacy security weakness of the encryption scheme based on random permutation, we propose an encryption scheme based on subtly designed invertible matrix which consists of both multiplication and addition perturbations. Specifically, our main contributions can be summarized as follows:

\begin{itemize}
\item We propose a novel encryption scheme for matrix multiplication based on subtly designed invertible matrix, which could protect the input original data as well as the output result of the outsourced task from leakage. As a consequence, the cloud server is restricted from learning any additional useful information about the clients' sensitive data, such as the statistic information of zero elements.

\item We show that our proposed encryption scheme can be applied to, except for the outsourced MMC task, other kinds of outsourced tasks. We take typical machine learning algorithms linear regression (LR) and principal component analysis (PCA) as examples and present concrete outsourcing LR and PCA protocol. The resulting schemes provide guarantees of privacy-protection to the original data.

\item The theoretical performance analysis indicates that our encryption scheme introduces no extra communication overhead and has the same computation complexity as the state-of-the-art scheme by exploring optimized matrix-chain multiplication method for our subtly designed invertible matrix. We further give a detailed implementation of the proposed protocol built with C++ and simulation results show that the efficiency of outsourcing protocol based on our proposed encryption scheme is comparable to that of the state-of-the-art scheme.
\end{itemize}

\subsection{Orgnization}
The remainder of this paper proceeds as follows. Section \ref{sec:2} introduces the problem statement of outsourcing tasks to cloud and some essential preliminaries. In Section \ref{sec:3}, we review the random-permutation-based encryption scheme proposed by Lei et al. \cite{lei2014achieving} and describe the inherent weakness of their work pointed out by \cite{zhao2018sparse,zhang2019practical,fu2017secure,wang2018more,kumar2019secure}. To solve this flaw, we propose a novel encryption scheme based on the subtly designed invertible matrix in Section \ref{sec:4}. Based on our proposed encryption scheme, we further give out outsourcing LR and PCA protocols to present the applicability of our proposed encryption scheme in the next section. And then Section \ref{sec:6} gives the theoretical analysis about privacy security, verifiability and efficiency of these protocols, followed by Section \ref{sec:7} which presents the performance evaluation by experiments. Finally, some conclusions are drawn in Section \ref{sec:8}.

\section{Problem statement and preliminaries}\label{sec:2}

In this section, we first introduce the system and threat model where the resource-constrained client outsources a task concerned with sensitive data to the cloud server. And for this typical model, design goals of the proposed encryption scheme are listed. Then we give a description about higher privacy security definition and mathematical background of random permutation.

\subsection{System and threat model}

\begin{figure}[ht]
\centering
\includegraphics[width=9cm]{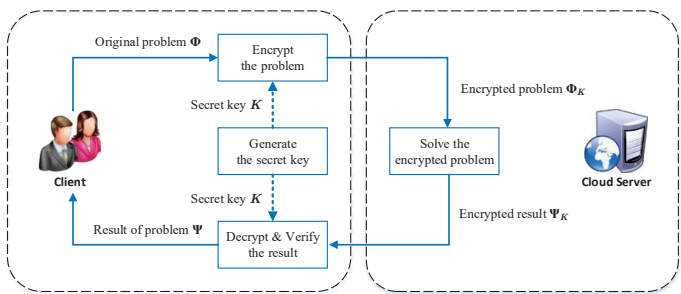}
\caption{System model for outsourcing task}
\label{sys_model}
\end{figure}

In this paper, we consider a typical situation involving a client and a cloud server, as shown in Fig. \ref{sys_model}. The client 
wishes to solve a large-scale computational problem $\bf{\Phi}$ and get the result $\bf{\Psi}$ of it, \textit{e.g.} large-scale matrix multiplication. However, it is generally a time-consuming process for a resource-constrained client to solve the problem $\bf{\Phi}$. So, outsourcing intensive tasks to the cloud becomes a promising method for resource-constrained clients. They can leverage computation resources of cloud servers to complete various
intensive tasks, typically in a flexible pay-per-use manner \cite{wang2011secure,xu2014proof}. 

While the cloud server may be malicious, which means the server could be curious-and-lazy \cite{wang2011secure,chen2014highly,chen2014new}.
On the one hand, the cloud may try to analyze the outsourced data of the problem $\bf{\Phi}$ to learn additional information. As the outsourced data and the result $\bf{\Psi}$ may contain sensitive information such as traffic data and genomic data, leakage of this information to malicious attackers will endanger the safety of society and individual. To protect this sensitive information, the client is supposed to encrypt the data $\bf{\Phi}$ using his/her secret key $K$ before outsourcing, and the cloud solves this problem over encrypted outsourced data ${\bf{\Phi}}_K$. The result ${\bf{\Psi}}_K$ got by the cloud is also encrypted by $K$. Only using client's secret key $K$, can the encrypted data ${\bf{\Phi}}_K$ and ${\bf{\Psi}}_K$ be decrypted.
On the other hand, the operation details inside the cloud server are not transparent enough to the client. As a result, the cloud server may be motivated by huge financial incentives (\textit{e.g.} saving the computing resources) to behave unfaithfully and return incorrect or even random results. Besides, the returned result might also be false because of software bugs, hardware failures or outside attacks. Thus, result verification is necessary. After receiving ${\bf{\Psi}}_K$ and decrypt it, the client has to verify whether the result $\bf{\Psi}$ is correct.

\subsection{Design goals}

To enable secure and practical outsourcing of various intensive tasks under the aforementioned model, our encryption scheme design should achieve the following goals.

\begin{itemize}
\item \textbf{Correctness.} Any cloud server which executes the designed protocol strictly must produce a result that can be decrypted and verified successfully by the client. 
\item \textbf{Input/Output Privacy.} The cloud server could derive no sensitive and additional useful information of the input dataset ${\bf{\Phi}}$ and output result ${\bf{\Psi}}$ during solving the problem over encrypted outsourced data.
\item \textbf{Verifiability.} The client should be able to verify the correctness of the result returned by the cloud server so that any false result can be resisted.
\item \textbf{Efficiency.} The local computation overhead of the client should be significantly less than solving the original problem on his/her own. 
The computation burden of solving the encrypted problem on the cloud server is supposed to be comparable with that of solving the original one. 
Moreover, computation cost introduced by protecting privacy should also be as minimal as possible. 
\item \textbf{Applicability.} The proposed encryption scheme is expected to be applied to not only matrix multiplication but also other kinds of outsourced tasks.
\end{itemize}

\subsection{Higher privacy security definition}

As described in \cite{fu2017secure,kumar2019secure,wang2018more,zhao2018sparse,zhang2019practical}, encryption schemes based on matrix transformation such as random permutation may leak the statistic information of zero elements in the original data, and then the attacker may exploit this additional useful information to recover the encrypted sensitive data. We call it \textit{zero elements attack} (ZEA). Referring to the definition of computational indistinguishability mentioned in \cite{katz2014introduction,salinas2015efficient}, we give a formal privacy security definition for ZEA attack in \textbf{Definition 1}.

We first define ZEA indistinguishability (IND-ZEA) experiment for the matrix-transformation-based encryption scheme in Algorithm \ref{alg_ind_zea}.
\begin{algorithm}
\caption{IND-ZEA experiment $\textmd{ZEA}_{\mathcal{A}}(\mathcal{K})$}
\label{alg_ind_zea}
\begin{algorithmic}[1]
\STATE Given a security parameter $\mathcal{K}$, the client (\textit{i.e.} challenger $\mathcal{C}$ ) generates the secret key $sk$.

\STATE The attacker $\mathcal{A}$ is given oracle access to $Enc(sk,\cdot)$ and outputs a pair of matrices ${\bf{T}}_0$ and ${\bf{T}}_1$ of the same dimension to the client.

\STATE A random bit $b \leftarrow \{ 0,1\}$ is chosen uniformly by the client, and then a ciphertext $Enc(sk,{\bf{T}}_b)$ is computed and given to the attacker.

\STATE The attacker $\mathcal{A}$ continues to have oracle access to $Enc(sk,\cdot)$. $\mathcal{A}$ outputs a bit $b'$ relying on the statistic information of zero elements in $Enc(sk,{\bf{T}}_b)$.

\STATE If $b=b'$, the output of the experiment is $1$; otherwise, $0$.
\end{algorithmic}
\end{algorithm}

\noindent \textbf{Definition 1.} We say that a matrix-transformation-based encryption scheme is IND-ZEA, if there exists a negligible function $\textbf{negl}$ such that, for all polynomial-time attackers $\mathcal{A}$, the probability
$$\left| \Pr (\textmd{ZEA}_{\mathcal{A}}(\mathcal{K}) = 1) - \frac{1}{2} \right| \le negl(\mathcal{K})$$

\subsection{Mathematical background of random permutation}

\noindent \textbf{Permutation function} $\pi$ rearranges all the elements of a set into another order, which has been well studied in group theory and combinatorics. This function can be viewed as a bijection function from a set $\bf{S}$ to itself. In Cauchy’s two-line notation, there are preimage elements in the first row (\textit{i.e.} the original set $\bf{S}$), and the corresponding image elements are listed in the second row (\textit{i.e.} the reordering set ${\bf{S}}'$). So, the permutation function can be expressed as
\begin{align}
\pi :\left( 
  {\begin{array}{*{20}{c}}
    s_1 & s_2 & \cdots &s_n \\
    {s_1'} & {s_2'} & \cdots & {s_n'}
  \end{array}} 
\right)
\end{align}
where $\pi(s_i) = s_i'(i=1,\cdots,n)$, and $\pi ^ {-1}$ denotes the inverse function of $\pi$. Here, $|{\bf{S}}|=n$, and $s_1,s_2,\cdots,s_n$ are the $n$ different elements of the set $|{\bf{S}}|$, so are the $s_1',s_2',\cdots,s_n'$. The random permutation can be generated by Algorithm \ref{alg_RPG}.


\begin{algorithm}
\caption{Random permutation generation}
\label{alg_RPG}
\begin{algorithmic}[1]
\REQUIRE A number $n$ representing the size of set $\bf{S}$
\ENSURE A random permutation $\pi$

\STATE {Construct a set ${\bf{S}}=\{1, 2, \cdots, n\}$ (\textit{i.e.} $s_i = i$).} 

\FOR{$i \in [2,n]$}
\STATE{Select a random integer $j$ where $1 \le j \le i$.}
\STATE{Swap $s_i$ and $s_j$.}
\ENDFOR
\STATE{Then a reodered set ${\bf{S}}'$ is constructed and set $\pi(s_i) = s_i'$, where $i=1,\cdots,n$}

\end{algorithmic}
\end{algorithm}

Algorithm \ref{alg_RPG} is due to Durstenfeld \cite{durstenfeld1964algorithm}, it is generally called Fisher-Yates shuffle \cite{knuth1997art}. There are several variants to generate a random permutation but the asymptotic time complexity of Algorithm \ref{alg_RPG} has already been optimal.

\noindent \textbf{Kronecker delta function} $\delta _{x,y}$ has two variables. The output of this function is $1$ if the variables are equal, and $0$ otherwise, which can be defined as
\begin{align}
{\delta _{x,y}} = \left\{ 
  {\begin{array}{*{20}{c}}
    {1,}&{x = y,}\\
    {0,}&{x \ne y.}
  \end{array}} \right.
\end{align}

\section{Review of encryption scheme based on random permutation}\label{sec:3}

In this section, we first review the encryption scheme \cite{lei2014achieving} based on random permutation. And then, we present the security weakness pointed out by \cite{zhao2018sparse,zhang2019practical,fu2017secure,wang2018more,kumar2019secure}.

\subsection{Review of Lei et al.'s encryption scheme}

Before outsourcing the multiplication between two large matrices ${\bf{X}} \in {\mathbb{R}}^{m \times n}$ and ${\bf{Y}} \in {\mathbb{R}}^{n \times s}$ to the malicious cloud, the resource-constrained client has to encrypt $\bf{X}$ and $\bf{Y}$ to protect the sensitive data from disclosure. The encryption scheme presented by Lei et al. contains five phases, namely secret key generation, MMC encryption, MMC in the cloud, MMC decryption, Result verification.

\subsubsection{Secret key generation} 
Here we describe the secret key generation as shown in Algorithm \ref{alg_KG}. Note that the secret key ${\bf{P}}_1$, ${\bf{P}}_2$ and ${\bf{P}}_3$ can be used only one time and thus have to be generated each time for different MMC tasks.

\begin{algorithm}
\caption{Secret key generation}
\label{alg_KG}
\begin{algorithmic}[1]
\REQUIRE A security parameter $\mathcal{K}$ and the dimension $m, n, s$
\ENSURE
Secret key $K$: ${\bf{P}}_1, {\bf{P}}_2, {\bf{P}}_3$

\STATE {Given a security parameter $\mathcal{K}$, which specifies a key space $\mathcal{P}_\alpha,\mathcal{P}_\beta,\mathcal{P}_\gamma$ where $0 \not\in \mathcal{P}_\alpha \cup \mathcal{P}_\beta \cup \mathcal{P}_\gamma$, the client picks sets of random numbers $\{ \alpha_1, \cdots, \alpha_m \} \leftarrow \mathcal{P}_\alpha$, $\{ \beta_1, \cdots, \beta_n \} \leftarrow \mathcal{P}_\beta$, $\{ \gamma_1, \cdots, \gamma_s \} \leftarrow \mathcal{P}_\gamma$.} 

\STATE{The client invokes the Algorithm \ref{alg_RPG} given the dimension $m, n, s$ and get the random permutation $\pi_1,\pi_2,\pi_3$.}

\STATE{Then the client generates matrices ${\bf{P}}_1, {\bf{P}}_2, {\bf{P}}_3$, where ${\bf{P}}_1(i,j)=\alpha_i \delta_{\pi_1(i),j}, {\bf{P}}_2(i,j)=\beta_i \delta_{\pi_2(i),j}, {\bf{P}}_3(i,j)=\gamma_i \delta_{\pi_3(i),j}$.}

\STATE{Meanwhile, the corresponding inverse matrices of ${\bf{P}}_1, {\bf{P}}_2, {\bf{P}}_3$ are ${\bf{P}}_1^{-1}(i,j)=(\alpha_j)^{-1} \delta_{\pi_1^{-1}(i),j}, {\bf{P}}_2^{-1}(i,j)=(\beta_j)^{-1} \delta_{\pi_2^{-1}(i),j}, {\bf{P}}_3^{-1}(i,j)=(\gamma_j)^{-1} \delta_{\pi_3^{-1}(i),j}$.}

\end{algorithmic}
\end{algorithm}

\subsubsection{MMC encryption} After generating the secret key ${\bf{P}}_1, {\bf{P}}_2, {\bf{P}}_3$, the client encrypts $\bf{X}$ and $\bf{Y}$ by computing ${\bf{X}}' = {\bf{P}}_1{\bf{X}}{\bf{P}}_2^{-1}$ and ${\bf{Y}}' = {\bf{P}}_2{\bf{Y}}{\bf{P}}_3^{-1}$. The computational process can be expressed as follows:

\begin{align}
{\bf{X}} = 
\left[ 
{\begin{array}{*{20}{c}}
{x_{11}}& \cdots & x_{1n}\\
 \vdots & \ddots & \vdots \\
x_{n1}& \cdots &{x_{nn}}
\end{array}} 
\right]
\end{align}

\begin{align}
{\bf{P}}_1{\bf{X}} = 
\left[ 
{\begin{array}{*{20}{c}}
{\alpha_1 x_{\pi_1(1),1}} & \cdots & {\alpha_1 x_{\pi_1(1),n}}\\
 \vdots & \ddots & \vdots \\
{\alpha_i x_{\pi_1(i),1}} & \cdots & {\alpha_i x_{\pi_1(i),n}} \\
 \vdots & \ddots & \vdots \\
{\alpha_n x_{\pi_1(n),1}}& \cdots & {\alpha_n x_{\pi_1(n),n}}
\end{array}} 
\right]
\end{align}

\begin{align}
\label{Lei_X_enc}
\begin{array}{l}
{\bf{X}}' = {\bf{P}}_1{\bf{X}}{\bf{P}}_2^{-1} = \\
\left[ 
\setlength{\arraycolsep}{0.39pt}
{\begin{array}{*{20}{c}}
{\frac{\alpha_1}{\beta_1} x_{\pi_1(1),\pi_2(1)}} & \cdots & {\frac{\alpha_1}{\beta_j} x_{\pi_1(1),\pi_2(j)}} & \cdots & {\frac{\alpha_1}{\beta_n} x_{\pi_1(1),\pi_2(n)}}\\
 \vdots & \ddots & \vdots & \ddots & \vdots \\
{\frac{\alpha_i}{\beta_1} x_{\pi_1(i),\pi_2(1)}} & \cdots & {\frac{\alpha_i}{\beta_j} x_{\pi_1(i),\pi_2(j)}} & \cdots & {\frac{\alpha_i}{\beta_n} x_{\pi_1(i),\pi_2(n)}}\\
 \vdots & \ddots & \vdots & \ddots & \vdots \\
{\frac{\alpha_n}{\beta_1} x_{\pi_1(n),\pi_2(1)}}& \cdots & {\frac{\alpha_n}{\beta_j} x_{\pi_1(n),\pi_2(j)}} & \cdots & {\frac{\alpha_n}{\beta_n} x_{\pi_1(n),\pi_2(n)}}
\end{array}} 
\right]
\end{array}
\end{align}
The process of computing ${\bf{Y}}' = {\bf{P}}_2{\bf{Y}}{\bf{P}}_3^{-1}$ is similar, which can be expressed as follows:
\begin{align}
{\bf{Y}}'(i,j) = {\frac{\beta_i}{\gamma_j} y_{\pi_2(i),\pi_3(j)}}
\end{align}
Later, the encrypted problem ${\bf{\Phi}}_K({\bf{X}}',{\bf{Y}}')$ will be outsourced to the cloud.

\subsubsection{MMC in the cloud} After receiving the encrypted task ${\bf{\Phi}}_K({\bf{X}}',{\bf{Y}}')$, the cloud computes ${\bf{Z}}' = {\bf{X}}' {\bf{Y}}'$ by any MMC method and then returns the encrypted result ${\bf{Z}}'$ to the client.

\subsubsection{MMC decryption} Upon receipt of the returned encrypted result ${\bf{Z}}'$, the client decrypts it by computing ${\bf{Z}} = {\bf{P}}_1^{-1} {\bf{Z}}' {\bf{P}}_3$ to get the real result ${\bf{Z}}$, where 
\begin{align}
{\bf{Z}}(i,j) = {\frac{\gamma_{\pi_3^{-1}(j)}}{\alpha_{\pi_1^{-1}(i)}} z'_{\pi_1^{-1}(i),\pi_3^{-1}(j)}}
\end{align}

\subsubsection{Result verification} The client generates a $s \times 1$ random $0/1$ vector $\bf{r}$ and computes ${\bf{P}} = {\bf{X}} \times ({\bf{Yr}}) - {\bf{Zr}}$. If ${\bf{P}} \neq (0, \cdots, 0)$, the client refuses the result ${\bf{Z}}$. To make the verification more robust, the client may repeat this process in a loop for 
$l (l \ll m,n,s)$
times. The result ${\bf{Z}}$ won't be accepted
until passing all the verification. This method is well known as Monte Carlo verification algorithm \cite{motwani1995randomized}.

\subsection{Security weakness analysis}

The encryption scheme based on random permutation is efficient, however it suffers from some weakness pointed out by \cite{zhao2018sparse,zhang2019practical,fu2017secure,wang2018more,kumar2019secure} in the sense of security requirement. As mentioned above, the client encrypts the original data $\bf{\Phi}$ in order that the cloud can't learn any additional useful information from the encrypted data ${\bf{\Phi}}_K$, nevertheless the encryption scheme based on random permutation leaks the statistic information of zeros in the original data $\bf{\Phi}$.

As shown in equation (\ref{Lei_X_enc}), ${\bf{X}}'(i,j) = \frac{\alpha_i}{\beta_j} x_{\pi_1(i),\pi_2(j)}$. If $x_{\pi_1(i),\pi_2(j)} = 0$, then ${\bf{X}}'(i,j) = 0$. A numerical example was given in \cite{fu2017secure} to illustrate the inherent weakness: assume that ${\bf{X}} = 
\begin{bmatrix}
1 & 0 & 2 \\
0 & 3 & 4
\end{bmatrix} 
$, the client generates the secret key $\{\alpha_1, \alpha_2\} = \{1, 2\}$, $\{\beta_1, \beta_2, \beta_3\} = \{3,4,5\}$, $\pi_1 =
\begin{pmatrix}
1 & 2 \\
2 & 1
\end{pmatrix} 
$, $\pi_2 = 
\begin{pmatrix}
1 & 2 & 3 \\
3 & 1 & 2
\end{pmatrix} 
$, and then the client gets ${\bf{P}}_1 = 
\begin{bmatrix}
0 & 1 \\
2 & 0
\end{bmatrix} 
$, ${\bf{P}}_2 = 
\begin{bmatrix}
0 & 0 & 3 \\
4 & 0 & 0 \\
0 & 5 & 0
\end{bmatrix}$, ${\bf{P}}_2^{-1} = 
\begin{bmatrix}
0 & \frac{1}{4} & 0 \\
0 & 0 & \frac{1}{5} \\
\frac{1}{3} & 0 & 0
\end{bmatrix}$. Consequently, ${\bf{X}}$ is encrypted as ${\bf{X}}' = {\bf{P}}_1 {\bf{X}} {\bf{P}}_2 = 
\begin{bmatrix}
\frac{4}{3} & 0 & \frac{3}{5} \\
\frac{4}{3} & \frac{1}{2} & 0
\end{bmatrix}$ and the number of zero elements of ${\bf{X}}'$ is same as that of ${\bf{X}}$. This is also similar for the matrix ${\bf{Y}}$ and its ciphertext ${\bf{Y}}'$. 

It is proved in \cite{fu2017secure} that the privacy of MMC is not protected and can't satisfy IND-ZEA. Let matrices ${\bf{P}}_1,{\bf{P}}_2,{\bf{P}}'_1,{\bf{P}}'_2$ be the secret key generated by the client, where 
${\bf{P}}_1(i,j)=\alpha_i \delta_{\pi_1(i),j}$, ${\bf{P}}_2(i,j)=\beta_i \delta_{\pi_2(i),j}$, ${\bf{P}}'_1(i,j)=\alpha'_i \delta_{\pi'_1(i),j}$, ${\bf{P}}'_2(i,j)=\beta'_i \delta_{\pi'_2(i),j}$. Given two matrices ${\bf{R}}$ and ${\bf{Q}}$ whose elements are respectively $r_{ij}$ and $q_{ij}$ chosen by the adversary $\mathcal{A}$, the client encrypts these matrices as follows:
\begin{align}
\begin{split}
{\bf{R}}'(i,j) = \frac{\alpha_i}{\beta_j} r_{\pi_1(i),\pi_2(j)} \\
{\bf{Q}}'(i,j) = \frac{\alpha'_i}{\beta'_j} q_{\pi'_1(i),\pi'_2(j)}
\end{split}
\end{align} 
According to the number of zero elements in matrices ${\bf{R}}$ and ${\bf{Q}}$, the adversary $\mathcal{A}$ can distinguish ${\bf{R}}'$ and ${\bf{Q}}'$.


\noindent \textbf{Analysis.} 
From the encryption scheme design perspective, this weakness results from only using multiplication perturbation without addition \cite{zhou2018efficiently}. So to solve this security weakness, we propose a more sophisticated scheme employing both multiplicative and additive perturbation, meanwhile maintaining the comparable efficiency to \cite{lei2014achieving} by exploiting the optimized matrix-chain multiplication, which will be described in the computation complexity analysis of Section \ref{sec:6} in detail. 

\section{The proposed encryption scheme}\label{sec:4}


In this section, we first introduce how to construct the invertible matrix based on the lemma presented in \cite{miller1981inverse}. 
Then we give the necessary condition to assure the designed matrix is invertible and prove it. 
Finally, we describe our proposed scheme on the basis of this subtly designed invertible matrix and the optimized matrix-chain multiplication.

\subsection{Subtly designed invertible matrix}

\noindent \textbf{Lemma 1} (\cite{miller1981inverse})\textbf{.}  ${\bf{G}}$ is an arbitrary $n \times n$ nonsingular square matrix, and ${\bf{H}}$ is a rank 1 square matrix of the same dimension. If ${\bf{G}}+{\bf{H}}$ is nonsingular, then the inverse of ${\bf{G}}+{\bf{H}}$ can be represented as follows:
\begin{align}
({\bf{G}} + {\bf{H}})^{-1} = {\bf{G}}^{-1} - \frac{{\bf{G}}^{-1} {\bf{H}} {\bf{G}}^{-1}}{1+tr({\bf{H}} {\bf{G}}^{-1})} 
\end{align}
where $tr({\bf{H}} {\bf{G}}^{-1})$ is the trace of ${\bf{H}} {\bf{G}}^{-1}$.

In virtue of Lemma 1, we use the sum of a nonsingular square matrix and a rank 1 square matrix of the same dimension to construct the invertible matrix. As the random-permutation-based matrix ${\bf{P}}$ is invertible which implies it is also nonsingular, we apply ${\bf{P}}$ as the nonsingular square matrix. For the rank 1 square matrix ${\bf{H}}$, we employ a matrix as follows:
\begin{equation} 
\begin{split}
{\bf{H}} &= \left[h_1, h_2, \cdots ,h_n \right]^T \left[1, 1, \cdots, 1 \right] \\
& = \left[ 
{\begin{array}{*{20}{c}}
{{h_1}}&{{h_1}}& \cdots &{{h_1}}\\
{{h_2}}&{{h_2}}& \cdots &{{h_2}}\\
 \vdots & \vdots & \ddots & \vdots \\
{{h_n}}&{{h_n}}& \cdots &{{h_n}}
\end{array}} \right]
\end{split}
\end{equation}
where $h_i \neq 0 (i \in [1,n])$ is a random number. 
So the subtly designed matrix can be represented as ${\bf{M}} = {\bf{P}} + {\bf{H}}$, consisting both multiplicative and additive perturbation.
Furthermore, we need to prove how to guarantee that ${\bf{P}} + {\bf{H}}$ is nonsingular (\textit{i.e.} the absolute value of determinant $|| {\bf{P}} + {\bf{H}} || \neq 0$).

\subsection{The necessary condition and proof}
Prior to giving the condition required to be guaranteed, we first introduce a lemma which will be used in the proof.

\noindent \textbf{Lemma 2} (\cite{zhang2017matrix})\textbf{.} Suppose that ${\bf{A}},{\bf{B}},{\bf{C}},{\bf{D}}$ are matrices of dimension $a \times a$, $a \times b$, $b \times a$, and $b \times b$, respectively. If ${\bf{A}}$ is invertible, the determinant of the $(a+b) \times (a+b)$ matrix 
$\begin{bmatrix}
{\bf{A}} & {\bf{B}}\\
{\bf{C}} & {\bf{D}} 
\end{bmatrix}$ can be represented as

\begin{equation}
\label{lemma2}
{\begin{vmatrix}
{\bf{A}} & {\bf{B}}\\
{\bf{C}} & {\bf{D}} 
\end{vmatrix}}= |{\bf{A}}| \times |{\bf{D}} - {\bf{C}}{{\bf{A}}^{-1}}{\bf{B}}|
\end{equation}

Then, the necessary condition is given in Theorem 1, followed by the proof.

\noindent \textbf{Theorem 1.} Only when $p_i \neq 0$, $h_i \neq 0$ and $\sum\limits_{i = 1}^n {\frac{h_i}{p_i}} \neq -1$, is $|| {\bf{P}} + {\bf{H}} || \neq 0$ guaranteed, namely the subtly designed matrix ${\bf{M}} = {\bf{P}} + {\bf{H}}$ is nonsingular or invertible.

\noindent \textbf{Proof.} Since the \textit{column switching} only changes the sign of the determinant of a matrix, the absolute value of this determinant won't be changed. We use ${\bf{P}}_{diag}$ to represent the diagonal matrix generated by switching columns of ${\bf{P}}$. As all columns of ${\bf{H}}$ are same (\textit{i.e.} $[h_1, h_2, \cdots, h_n]^T$), column switching won't change the rank 1 matrix ${\bf{H}}$. Therefore we can get 
\begin{align}
\label{row_switch}
|| {\bf{P}} + {\bf{H}} || = || {\bf{P}}_{diag} + {\bf{H}} || 
\end{align}
Take 
${\bf{P}} = \begin{bmatrix}
0 & 0 & 1 \\
0 & 2 & 0 \\
3 & 0 & 0 \\
\end{bmatrix}$ and 
${\bf{H}} = \begin{bmatrix}
4 & 4 & 4 \\
5 & 5 & 5 \\
6 & 6 & 6 \\
\end{bmatrix}$ as a toy example, $|| {\bf{P}} + {\bf{H}} || = 
\begin{Vmatrix}
4 & 4 & 5 \\
5 & 7 & 5 \\
9 & 6 & 6 \\
\end{Vmatrix} = |-57| = 57$ and $|| {\bf{P}}_{diag} + {\bf{H}} || = 
\begin{Vmatrix}
5 & 4 & 4 \\
5 & 7 & 5 \\
6 & 6 & 9 \\
\end{Vmatrix} = |57| = 57$, then $|| {\bf{P}} + {\bf{H}} || = || {\bf{P}}_{diag} + {\bf{H}} ||$.

Therefore, we are supposed to ensure $|| {\bf{P}}_{diag} + {\bf{H}} || \neq 0$ on account of the equation (\ref{row_switch}). The element of ${\bf{P}}_{diag}$ is denoted as $p_i, i \in [1,n]$, \textit{i.e.} ${\bf{P}}_{diag} = diag(p_1, p_2, \cdots, p_n)$. Hence the ${\bf{P}}_{diag} + {\bf{H}}$ can be represented as follows:

\begin{equation} 
\begin{split}
{\bf{P}}_{diag} + {\bf{H}} = \left[ 
{\begin{array}{*{20}{c}}
{p_1 + h_1} & {{h_1}}& \cdots &{{h_1}}\\
{{h_2}}&{p_2 + h_2}& \cdots &{{h_2}}\\
 \vdots & \vdots & \ddots & \vdots \\
{{h_n}}&{{h_n}}& \cdots &{p_n + h_n}
\end{array}} \right]
\end{split}
\end{equation} 

Except for column switching, other kinds of elementary operations such as row multiplication and row addition won't change whether the determinant of a matrix is $0$. As a result, the condition that $|| {\bf{P}}_{diag} + {\bf{H}} ||$ doesn't equal $0$ after row multiplication and row addition is equivalent to $|| {\bf{P}}_{diag} + {\bf{H}} || \neq 0$. 

\textit{Row multiplication} means each element in a row can be multiplied by a non-zero constant and \textit{row addition} implies a row can be replaced by the sum of that row and a multiple of another row. We use $row_i$ to denote the $i$-th row of ${\bf{P}}_{diag} + {\bf{H}}$ and then perform the row multiplication $row_i \times \prod {{h_j}}, (i \in [1,n], j \in [1,n], j \neq i)$ and row addition $row_i - row_n, (i \in [1,n-1])$ as follows:

\mathleft
\begin{equation} 
\begin{array}{l}
{\bf{P}}_{diag} + {\bf{H}} \xrightarrow[\hfill]{row_i \times \prod\limits_{j = 1, j \neq i}^n {{h_j}}} ({\bf{P}}_{diag} + {\bf{H}})' = \\
\begin{bsmallmatrix}
{{p_1}\prod\limits_{\mathclap{j = 2}}^n {{h_j}}  + \prod\limits_{j = 1}^n {{h_j}} }& {\cdots} &{\prod\limits_{j = 1}^n {{h_j}} }& {\cdots} &{\prod\limits_{j = 1}^n {{h_j}} }\\
 \vdots & \ddots &\vdots & \ddots & \vdots \\
{\prod\limits_{j = 1}^n {{h_j}} } & \cdots &{{p_i}\prod\limits_{ \mathclap{j = 1} \atop \mathclap{j \neq i}}^n {{h_j}}  + \prod\limits_{j = 1}^n {{h_j}} }& \cdots &{\prod\limits_{j = 1}^n {{h_j}} }\\
\vdots & \ddots &\vdots & \ddots & \vdots \\
{\prod\limits_{j = 1}^n {{h_j}} }& \cdots &{\prod\limits_{j = 1}^n {{h_j}} }& \cdots &{{p_n}\prod\limits_{j = 1}^{\mathclap{n-1}} {{h_j}}  + \prod\limits_{j = 1}^n {{h_j}} }
\end{bsmallmatrix} 
\end{array}
\end{equation} 
\mathcenter

\mathleft
\begin{equation} 
\begin{array}{l}
({\bf{P}}_{diag} + {\bf{H}})' \xrightarrow[\hfill]{row_i - row_n,(i \neq n)} ({\bf{P}}_{diag} + {\bf{H}})'' = \\
\begin{bsmallmatrix}
{p_1}\prod\limits_{\mathclap{j = 2}}^n {{h_j}} & 0 & 0 & \cdots & 0 & { - {p_n}\prod\limits_{j = 1}^{n{\rm{ - }}1} {{h_j}} }\\
0 & \ddots & \ddots & 0 & 0 & { - {p_n}\prod\limits_{j = 1}^{n{\rm{ - }}1} {{h_j}} }\\
0 & \ddots & {{p_i}\prod\limits_{j = 1 \atop j \ne i}^n {{h_j}} } & \ddots & \vdots & \vdots\\
\vdots & \ddots & \ddots & \ddots & 0 & { - {p_n}\prod\limits_{j = 1}^{n{\rm{ - }}1} {{h_j}} } \\
0 & 0 & \cdots & 0 & {{p_{n-1}}\prod\limits_{\mathclap{j = 1 \atop j \ne n-1}}^n {{h_j}} } & { - {p_n}\prod\limits_{j = 1}^{n{\rm{ - }}1} {{h_j}} }\\
{\prod\limits_{j = 1}^n {{h_j}} } & \cdots & {\prod\limits_{j = 1}^n {{h_j}} } & \cdots  &{\prod\limits_{j = 1}^n {{h_j}} } & {{p_n}\prod\limits_{j = 1}^{\mathclap{n-1}} {{h_j}}  + \prod\limits_{j = 1}^n {{h_j}} }
\end{bsmallmatrix}
\end{array}
\end{equation} 
\mathcenter

For the sake of simplifying the determinant computation of $({\bf{P}}_{diag} + {\bf{H}})''$, it can be divided into four block matrices in the form of 
$\begin{bmatrix}
{\bf{A}} & {\bf{B}}\\
{\bf{C}} & {\bf{D}} 
\end{bmatrix}$:
\begin{align*}
\begin{array}{l} 
{\bf{A}}_{(n-1)\times(n-1)} = diag({p_1}\prod\limits_{\mathclap{j = 2}}^n {{h_j}}, \cdots, {p_i}\prod\limits_{j = 1 \atop j \neq i}^n {{h_j}}, \cdots,{{p_{n-1}}\prod\limits_{\mathclap{j = 1 \atop j \ne n-1}}^n {{h_j}} }) \\
{\bf{B}}_{(n-1)\times 1} = [{ - {p_n}\prod\limits_{j = 1}^{n{\rm{ - }}1} {{h_j}} }, \cdots, { - {p_n}\prod\limits_{j = 1}^{n{\rm{ - }}1} {{h_j}} }] ^T \\
{\bf{C}}_{1\times(n-1)} = [{\prod\limits_{j = 1}^n {{h_j}}}, \cdots, {\prod\limits_{j = 1}^n {{h_j}}}] \\
{\bf{D}}_{1 \times 1} = [{{p_n}\prod\limits_{j = 1}^{\mathclap{n-1}} {{h_j}}  + \prod\limits_{j = 1}^n {{h_j}}}] \\
\end{array}
\end{align*}
${\bf{A}}$ is invertible since it is a diagonal matrix. Therefore according to Lemma 2, the determinant of $({\bf{P}}_{diag} + {\bf{H}})''$ can be worked out more easily as follows:
\mathleft
\begin{align}
\begin{split}
\label{eq_det}
&\quad |({\bf{P}}_{diag} + {\bf{H}})''| = |{\bf{A}}| \times |{\bf{D}} - {\bf{C}}{{\bf{A}}^{-1}}{\bf{B}}| \\
&= (\prod\limits_{i = 1}^{n - 1} {{p_i}} ){(\prod\limits_{i = 1}^n {{h_i}} )^{n - 2}}{h_n} \\
&\qquad \times [({p_n}\prod\limits_{i = 1}^{n-1} {{h_i}}  + \prod\limits_{i = 1}^n {{h_i}) - ( - {p_n}\prod\limits_{i = 1}^{n - 1} {{h_i}} )\sum\limits_{i = 1}^{n - 1} {\frac{{{h_i}}}{{{p_i}}}} ]} \\
&= (\prod\limits_{i = 1}^{n - 1} {{p_i}} ){(\prod\limits_{i = 1}^n {{h_i}} )^{n - 2}}\prod\limits_{i = 1}^n {{h_i}}  \times ({p_n} + {h_n} + {p_n}\sum\limits_{i = 1}^{n - 1} {\frac{{{h_i}}}{{{p_i}}}} )\\
&= (\prod\limits_{i = 1}^n {{p_i}} ){(\prod\limits_{i = 1}^n {{h_i}} )^{n - 1}}(1 + \sum\limits_{i = 1}^n {\frac{h_i}{p_i}} )
\end{split}
\end{align} 
\mathcenter

As mentioned above, ${\bf{P}} + {\bf{H}}$ is transformed to $({\bf{P}}_{diag} + {\bf{H}})''$ after a series of elementary operations, and elementary operations won’t change whether the determinant of a matrix is $0$. So from the equation (\ref{eq_det}) we can see, only if $p_i \neq 0$, $h_i \neq 0$ and $\sum\limits_{i = 1}^n {\frac{h_i}{p_i}} \neq -1$ is $||{\bf{P}} + {\bf{H}}|| \neq 0$ guaranteed, \textit{i.e.} ${\bf{M}} = {\bf{P}} + {\bf{H}}$ is nonsingular or invertible. $\hfill \blacksquare$

Utilizing this subtly designed invertible matrix ${\bf{M}} = {\bf{P}} + {\bf{H}}$, we propose a new encryption scheme to protect the privacy of clients' sensitive data.

\subsection{Our proposed encryption scheme}

To protect the sensitive information of the two large matrices ${\bf{X}} \in {\mathbb{R}}^{m \times n}$ and ${\bf{Y}} \in {\mathbb{R}}^{n \times s}$ from disclosure, the resource-constrained client has to encrypt $\bf{X}$ and $\bf{Y}$, and then sends the encrypted matrices ${\bf{X}}'$ and ${\bf{Y}}'$ to the cloud to perform the encrypted MMC task ${\bf{Z}}' = {\bf{X}}' {\bf{Y}}'$. Upon receiving the result ciphertext ${\bf{Z}}'$ returned from the cloud, the client recovers the result ${\bf{Z}} = {\bf{X}} {\bf{Y}}$ from ${\bf{Z}}'$ and verify the result. These phases are described in the following.

\subsubsection{Secret key generation} The secret key is generated as shown in Algorithm \ref{alg_our_KG}. The secret key ${\bf{M}}_1$, ${\bf{M}}_2$ and ${\bf{M}}_3$ will also be used only one time for each MMC task. As described in the computation complexity analysis of Section \ref{sec:6}, this phase is extremely fast. 

\begin{algorithm}
\caption{Secret key generation}
\label{alg_our_KG}
\begin{algorithmic}[1]
\REQUIRE A security parameter $\mathcal{K}$ and the dimension $m, n, s$
\ENSURE
Secret key $K$: ${\bf{M}}_1, {\bf{M}}_2, {\bf{M}}_3$

\STATE {Given the input, the client invokes the Algorithm \ref{alg_KG} to generate matrices ${\bf{P}}_1, {\bf{P}}_2, {\bf{P}}_3$, where ${\bf{P}}_1(i,j)=\alpha_i \delta_{\pi_1(i),j}, {\bf{P}}_2(i,j)=\beta_i \delta_{\pi_2(i),j}, {\bf{P}}_3(i,j)=\gamma_i \delta_{\pi_3(i),j}$.}

\STATE {With the security parameter $\mathcal{K}$, which specifies a key space $\mathcal{H}^{\alpha},\mathcal{H}^{\beta},\mathcal{H}^{\gamma}$ where $0 \not\in \mathcal{H}^{\alpha} \cup \mathcal{H}^{\beta} \cup \mathcal{H}^{\gamma}$, the client selects sets of random numbers $\{ h^\alpha_1, \cdots, h^\alpha_m \} \leftarrow \mathcal{H}^\alpha$, $\{ h^\beta_1, \cdots, h^\beta_n \} \leftarrow \mathcal{H}^\beta$, $\{ h^\gamma_1, \cdots, h^\gamma_s \} \leftarrow \mathcal{H}^\gamma$. Note that conditions $\sum\limits_{i = 1}^m {\frac{h^\alpha_i}{\alpha_i}} \neq -1$, $\sum\limits_{i = 1}^n {\frac{h^\beta_i}{\beta_i}} \neq -1$ and $\sum\limits_{i = 1}^s {\frac{h^\gamma_i}{\gamma_i}} \neq -1$ are supposed to be guaranteed.} 

\STATE{Then, the rank $1$ matrices ${\bf{H}}_1, {\bf{H}}_2, {\bf{H}}_3$ are constructed, where ${\bf{H}}_1 = [ h^\alpha_1, \cdots, h^\alpha_m]^T [1, \cdots, 1]$, ${\bf{H}}_2 = [ h^\beta_1, \cdots, h^\beta_n]^T [1, \cdots, 1]$, ${\bf{H}}_3 = [ h^\gamma_1, \cdots, h^\gamma_s]^T [1, \cdots, 1]$. }

\STATE{Finally, the client generates matrices 
${\bf{M}}_1 = {\bf{P}}_1 + {\bf{H}}_1$, 
${\bf{M}}_2 = {\bf{P}}_2 + {\bf{H}}_2$, 
${\bf{M}}_3 = {\bf{P}}_3 + {\bf{H}}_3$} as secret key.

\STATE{Meanwhile, the corresponding inverse matrices of ${\bf{M}}_1$, ${\bf{M}}_2$, ${\bf{M}}_3$ are 
${\bf{M}}_1^{-1}={\bf{P}}^{-1}_1 - \frac{1}{1+tr({\bf{H}}_1 {\bf{P}}^{-1}_1)} {\bf{P}}^{-1}_1 {\bf{H}}_1 {\bf{P}}^{-1}_1$, 
${\bf{M}}_2^{-1}={\bf{P}}^{-1}_2 - \frac{1}{1+tr({\bf{H}}_2 {\bf{P}}^{-1}_2)} {\bf{P}}^{-1}_2 {\bf{H}}_2 {\bf{P}}^{-1}_2$, 
${\bf{M}}_3^{-1}={\bf{P}}^{-1}_3 - \frac{1}{1+tr({\bf{H}}_3 {\bf{P}}^{-1}_3)} {\bf{P}}^{-1}_3 {\bf{H}}_3 {\bf{P}}^{-1}_3$.}

\end{algorithmic}
\end{algorithm}

\subsubsection{MMC encryption} With the generated secret key ${\bf{M}}_1, {\bf{M}}_2, {\bf{M}}_3$, the client encrypts $\bf{X}$ and $\bf{Y}$ by computing ${\bf{X}}' = {\bf{M}}_1{\bf{X}}{\bf{M}}_2^{-1}$ and ${\bf{Y}}' = {\bf{M}}_2{\bf{Y}}{\bf{M}}_3^{-1}$. To improve the computation efficiency, the encryption process can be executed as follows:
\begin{align}
\label{X_enc}
\begin{split}
{\bf{X}}' & = {\bf{M}}_1{\bf{X}}{\bf{M}}_2^{-1} \\
& = ({\bf{P}}_1 + {\bf{H}}_1){\bf{X}}[{\bf{P}}^{-1}_2 - \frac{{\bf{P}}^{-1}_2 {\bf{H}}_2 {\bf{P}}^{-1}_2}{1+tr({\bf{H}}_2 {\bf{P}}^{-1}_2)} ] \\
& = {\bf{P}}_1 {\bf{X}} {\bf{P}}^{-1}_2 - \frac{{\bf{P}}_1 {\bf{X}} {\bf{P}}^{-1}_2 {\bf{H}}_2 {\bf{P}}^{-1}_2}{1+tr({\bf{H}}_2 {\bf{P}}^{-1}_2)}  \\
& \qquad+ {\bf{H}}_1 {\bf{X}} {\bf{P}}^{-1}_2 - \frac{{\bf{H}}_1 {\bf{X}} {\bf{P}}^{-1}_2 {\bf{H}}_2 {\bf{P}}^{-1}_2}{1+tr({\bf{H}}_2 {\bf{P}}^{-1}_2)} 
\end{split}
\end{align}

The encryption process of ${\bf{Y}}' = {\bf{M}}_2{\bf{Y}}{\bf{M}}_3^{-1}$ is similar, which can be expressed as follows:
\begin{align}
\begin{split}
&{\bf{Y}}' = {\bf{P}}_2 {\bf{Y}} {\bf{P}}^{-1}_3 - \frac{{\bf{P}}_2 {\bf{Y}} {\bf{P}}^{-1}_3 {\bf{H}}_3 {\bf{P}}^{-1}_3}{1+tr({\bf{H}}_3 {\bf{P}}^{-1}_3)}  \\
& \qquad\quad + {\bf{H}}_2 {\bf{Y}} {\bf{P}}^{-1}_3 - \frac{{\bf{H}}_2 {\bf{Y}} {\bf{P}}^{-1}_3 {\bf{H}}_3 {\bf{P}}^{-1}_3}{1+tr({\bf{H}}_3 {\bf{P}}^{-1}_3)} 
\end{split}
\end{align}
After this encryption process is finished, the encrypted problem ${\bf{\Phi}}_K({\bf{X}}',{\bf{Y}}')$ will be outsourced to the cloud.

\subsubsection{MMC in the cloud} Once receiving the encrypted problem ${\bf{\Phi}}_K({\bf{X}}',{\bf{Y}}')$, the cloud computes ${\bf{Z}}' = {\bf{X}}' {\bf{Y}}'$ using any MMC way and then sends the encrypted result ${\bf{Z}}'$ to the client.

\subsubsection{MMC decryption} After receiving the encrypted result ${\bf{Z}}'$, the client performs decryption ${\bf{Z}} = {\bf{M}}_1^{-1} {\bf{Z}}' {\bf{M}}_3$ to get the real result ${\bf{Z}}$. For higher performance, this process can be computed as follows:
\begin{align}
\begin{split}
{\bf{Z}} & = {\bf{M}}_1^{-1} {\bf{Z}}' {\bf{M}}_3 \\
& = [{\bf{P}}^{-1}_1 - \frac{ {\bf{P}}^{-1}_1 {\bf{H}}_1 {\bf{P}}^{-1}_1}{1+tr({\bf{H}}_1 {\bf{P}}^{-1}_1)}] {\bf{Z}}' ({\bf{P}}_3 + {\bf{H}}_3) \\
& = {\bf{P}}^{-1}_1 {\bf{Z}}' {\bf{P}}_3 - \frac{{\bf{P}}^{-1}_1 {\bf{H}}_1 {\bf{P}}^{-1}_1 {\bf{Z}}' {\bf{P}}_3}{1+tr({\bf{H}}_1 {\bf{P}}^{-1}_1)}\\
& \qquad +{\bf{P}}^{-1}_1 {\bf{Z}}' {\bf{H}}_3 - \frac{{\bf{P}}^{-1}_1 {\bf{H}}_1 {\bf{P}}^{-1}_1 {\bf{Z}}' {\bf{H}}_3}{1+tr({\bf{H}}_1 {\bf{P}}^{-1}_1)} 
\end{split}
\end{align}

\subsubsection{Result verification} The Monte Carlo verification algorithm \cite{motwani1995randomized} is also adopted. The client selects a $s \times 1$ dimension $0/1$ vector $\bf{r}$ randomly and computes ${\bf{R}} = {\bf{X}} \times ({\bf{Yr}}) - {\bf{Zr}}$. If ${\bf{R}} \neq (0, \cdots, 0)$, the client refuses the result ${\bf{Z}}$. The client can also repeat this process in a loop for $l (l \ll m,n,s)$ times to make the verification more robust. The result ${\bf{Z}}$ won't be accepted until all the verification succeeds.

\subsection{Optimized matrix-chain multiplication}
To maintain the efficiency of the client during encryption and decryption phases, we explore a optimized matrix-chain multiplication method for the subtly designed invertible matrix, of which the main idea is deciding the sequence of the matrix multiplications involved to find the most efficient way to multiply these matrices. Take a $n \times n$ matrix ${\bf{T}}$ as example, we consider using the secret key ${\bf{M}}$ to encrypt/decrypt it, \textit{i.e.} multiply ${\bf{T}}$ with ${\bf{M}}$ (or ${\bf{M}}^{-1}$). The multiplication case consists of ${\bf{M}} {\bf{T}}$, ${\bf{T}} {\bf{M}}$, ${\bf{M}}^{-1} {\bf{T}}$ and ${\bf{T}} {\bf{M}}^{-1}$. The matrix-chain multiplication method can be expressed as follows:
\begin{align*}
\begin{split}
{\bf{M}} {\bf{T}} & = ({\bf{P}} {\bf{T}}) + ({\bf{H}} {\bf{T}}) \\
{\bf{T}} {\bf{M}} & = ({\bf{T}} {\bf{P}}) + ({\bf{T}} {\bf{H}}) \\
{\bf{M}}^{-1} {\bf{T}} & = ({\bf{P}}^{-1} {\bf{T}}) - \frac{{\bf{P}}^{-1} [{\bf{H}} ({\bf{P}}^{-1} {\bf{T}})]}{1+tr({\bf{H}} {\bf{P}}^{-1})}  \\
{\bf{T}} {\bf{M}}^{-1} & = ({\bf{T}}{\bf{P}}^{-1}) - \frac{[({\bf{T}} {\bf{P}}^{-1}) {\bf{H}}] {\bf{P}}^{-1}}{1+tr({\bf{H}} {\bf{P}}^{-1})}   \\
\end{split}
\end{align*}
Parentheses represent the priority order of multiplication, and the detailed computation complexity will be discussed in Section \ref{sec:6}.

\section{Other applications based on the proposed encryption scheme}\label{sec:5}

To present the applicability of our encryption scheme, we take typical machine learning algorithm \textit{linear regression} (LR) and \textit{principal component analysis} (PCA) as examples and then put forward the outsourcing protocol based on our proposed encryption scheme.

\subsection{LR based on our encryption scheme}

In the scientific and engineering community, \textit{linear regression} (LR) is a common machine learning approach to modeling the relationship between a dependant variable and independent variables, and widely used in many applications \cite{weisberg2005applied}. 
However, using a large-scale dataset generated in real-world settings to construct a LR model will lead to huge computation beyond client's computation power (\textit{e.g.} personal computer) \cite{weisberg2005applied,fan2014challenges}. So the resource-limited client may outsource the LR task to the cloud server. 
While the cloud server is not fully trusted by the client and generally curious-and-lazy, which means the cloud server will try to learn some additional information of the client's sensitive data and even return a false result. To deal with these issues, we present an outsourcing LR protocol based on our proposed encryption scheme. 

Prior to the protocol, we first give out the LR problem statement. The client has a large-scale dataset $[x_{i1}, x_{i2}, \cdots, x_{in}, y_i]_{i=1}^m$ of $m$ statistical units, each of which consists of $n$ independent variables ${\bf{x_i}} = [x_{i1}, x_{i2}, \cdots, x_{in}]$ and one dependent variable $y_i$. All independent and dependent variables are numerical and make up a $m \times n$ matrix ${\bf{X}} = [{\bf{x_1}}, {\bf{x_2}}, \cdots, {\bf{x_m}}]^T$ and a $m \times 1$ vector ${\bf{y}} = [y_1, y_2, \cdots, y_m]^T$. The client wishes to solve the LR problem ${\bf{\Phi}}({\bf{X}},{\bf{y}})$ to get the model ${\bf{y}} = {\bf{X}}\bm{\beta} + \beta_0$, where $\bm{\beta}$ is a $n \times 1$ vector called regression coefficient and $\beta_0$ is a scalar called intercept term. To determine a good estimation of $(\bm{\beta}, \beta_0)$, least squared error is the typical widely-used criterion expressed as
\begin{align}
\left\{ 
\begin{array}{l}
\bm{\beta} = (\bm{\mathcal{X}}^T \bm{\mathcal{X}})^{-1} \bm{\mathcal{X}}^T \bm{\mathcal{Y}}\\
\beta_0 = \bar{y} - \bar{\bf{x}}\bm{\beta}
\end{array} \right.
\end{align}
Here, 
$\bar{\bf{x}}$ is a $1 \times n$ vector $[\bar{x}_1, \bar{x}_2, \cdots, \bar{x}_n]$ where $\bar{x}_j = \frac{1}{m}\sum\limits_{i=1}^{m} x_{ij}, j \in [1,n]$, and $\bm{\mathcal{X}}$ is a $m \times n$ matrix whose $i$-th row is $\mathcal{X}_i = {\bf{x_i}} - \bar{\bf{x}} = [x_{i1}-\bar{x}_1, x_{i2}-\bar{x}_2, \cdots, x_{in}-\bar{x}_n]$. 
$\bar{y} = \frac{1}{m}\sum\limits_{i=1}^{m} y_i$ and $\bm{\mathcal{Y}}$ is a $m \times 1$ vector $[y_1-\bar{y}, y_2-\bar{y}, \cdots, y_m-\bar{y}]^T$. In addition, $m>n$ and $\bm{\mathcal{X}}^T \bm{\mathcal{X}}$ is assumed invertible.

Then, we describe the outsourcing LR protocol based on our proposed encryption scheme in five phases.

\subsubsection{Secret key generation}
Given a security parameter $\mathcal{K}$, the client first generates an invertible matrix ${\bf{M}} = {\bf{P}} + {\bf{H}}$ based on our proposed encryption scheme, where ${\bf{P}}(i,j)=p_i \delta_{\pi(i),j}$ and ${\bf{H}} = \left[h_1, \cdots ,h_n \right]^T \left[1, \cdots, 1 \right]$. 
Then the client selects a random number $k>0$ to construct a $m \times m$ diagonal matrix ${\bf{A}}$, the diagonal elements of which are randomly set $k$ or $-k$. 
Moreover, a random $n \times 1$ vector ${\bm{\mathcal{R}}} = [r_1, r_2, \cdots, r_n]^T$ is also chosen by the client to encrypt the sensitive data.

\subsubsection{LR encryption}
The client computes the $\bar{\bf{x}}$, $\bm{\mathcal{X}}$, $\bar{y}$, $\bm{\mathcal{Y}}$ locally according to the LR problem statement. And then, the client encrypts $\bm{\mathcal{X}}$ and $\bm{\mathcal{Y}}$ as shown in equation (\ref{LR_enc}).
\begin{align}
\label{LR_enc}
\left\{ 
\begin{array}{l}
\bm{\mathcal{X}}' = {\bf{A}} \bm{\mathcal{X}} {\bf{M}}\\
\bm{\mathcal{Y}}' = {\bf{A}} (\bm{\mathcal{Y}} + \bm{\mathcal{X}} {\bm{\mathcal{R}}} )
\end{array} \right.
\end{align}
After local computation, the encrypted LR task ${\bf{\Phi}}_K(\bm{\mathcal{X}}',\bm{\mathcal{Y}}')$ will be outsourced to the cloud server.

\subsubsection{LR in the cloud}
After receiving the outsourcing LR task ${\bf{\Phi}}_K(\bm{\mathcal{X}}',\bm{\mathcal{Y}}')$, the cloud solves the encrypted LR problem as follows:
\mathleft
\begin{align}
\begin{split}
\bm{\beta}' & = (\bm{\mathcal{X}}'^{T} \bm{\mathcal{X}}')^{-1} \bm{\mathcal{X}}'^{T} \bm{\mathcal{Y}}' \\
& = ({\bf{M}}^T \bm{\mathcal{X}}^T {\bf{A}}^T {\bf{A}} \bm{\mathcal{X}} {\bf{M}})^{-1} {\bf{M}}^T \bm{\mathcal{X}}^T {\bf{A}}^T {\bf{A}} (\bm{\mathcal{Y}} + \bm{\mathcal{X}} {\bm{\mathcal{R}}}) \\
& = {\bf{M}}^{-1} (\bm{\mathcal{X}}^T k^2 {\bf{I}} \bm{\mathcal{X}})^{-1} ({\bf{M}}^T)^{-1} {\bf{M}}^T \bm{\mathcal{X}}^T k^2 {\bf{I}} (\bm{\mathcal{Y}} + \bm{\mathcal{X}} {\bm{\mathcal{R}}}) \\
& = {\bf{M}}^{-1} (\bm{\mathcal{X}}^T \bm{\mathcal{X}})^{-1} \bm{\mathcal{X}}^T (\bm{\mathcal{Y}} + \bm{\mathcal{X}} {\bm{\mathcal{R}}}) \\
& = {\bf{M}}^{-1} [(\bm{\mathcal{X}}^T \bm{\mathcal{X}})^{-1} \bm{\mathcal{X}}^T \bm{\mathcal{Y}} + {\bm{\mathcal{R}}}] \\
& = {\bf{M}}^{-1}(\bm{\beta} + {\bm{\mathcal{R}}})
\end{split}
\end{align}
\mathcenter
where ${\bf{I}}$ denotes the identity matrix. And then, the cloud returns the encrypted result $\bm{\beta}'$ to the client.

\subsubsection{Result verification}
If the cloud computes the $\bm{\beta}'$ honestly, the equation $\bm{\beta}' = (\bm{\mathcal{X}}'^{T} \bm{\mathcal{X}}')^{-1} \bm{\mathcal{X}}'^{T} \bm{\mathcal{Y}}'$ holds. This equation is equivalent to following equation:
\begin{align}
\label{LR_verify}
\begin{split}
(\bm{\mathcal{X}}'^{T} \bm{\mathcal{X}}') \bm{\beta}' = \bm{\mathcal{X}}'^{T} \bm{\mathcal{Y}}' \\
\bm{\mathcal{X}}'^{T} (\bm{\mathcal{X}}' \bm{\beta}' - \bm{\mathcal{Y}}') = 0
\end{split}
\end{align}
Thus upon obtaining the result $\bm{\beta}'$, the client checks whether $\bm{\mathcal{X}}'^{T} (\bm{\mathcal{X}}' \bm{\beta}' - \bm{\mathcal{Y}}') = 0$ holds. If so, the client accepts the encrypted result $\bm{\beta}'$ and decrypts it; otherwise, claims the misbehavior of the cloud server.

\subsubsection{LR decryption} If the verification succeeds, the client decrypts the encrypted result $\bm{\beta}'$ to get the regression coefficient $\bm{\beta}$ and further computes the intercept term $\beta_0$ as follows:
\begin{align}
\left\{ 
\begin{array}{l}
\bm{\beta} = {\bf{M}} \bm{\beta}' - {\bm{\mathcal{R}}} \\
\beta_0 = \bar{y} - \bar{\bf{x}}\bm{\beta}
\end{array} \right.
\end{align}

\subsection{PCA based on our encryption scheme}

In the field of data analysis, with large datasets of many disciplines rapidly increasing, \textit{principal component analysis} (PCA) has become the most widely used technique for reducing the dimensionality of such datasets, increasing interpretability but at the same time minimizing information loss \cite{jolliffe2016principal}.
Nevertheless, performing PCA on large datasets generally leads to huge computation beyond client’s computing power, resulting in the resource-limited client resorting to cloud computing to complete the PCA task.
However, the cloud server is commonly curious-and-lazy as mentioned in Section \ref{sec:2}. Outsourcing the original data directly to the cloud server may cause leakage of privacy. Therefore, we put forward an outsourcing PCA protocol based on our proposed encryption scheme.

Before depicting the protocol, we first describe the PCA problem statement. 
Assume that the client has a large-scale dataset denoted as a $n \times m$ matrix ${\bf{D}} = [{\bf{x}}_1,{\bf{x}}_2,\cdots,{\bf{x}}_m]$, where ${\bf{x}}_i$ is a $n \times 1$ vector $[x_{i1},x_{i2}, \cdots ,x_{in}]^T$ representing the $i$-th column of ${\bf{D}}$. The covariance matrix of ${\bf{D}}$ can be computed by ${\bf{A}} = {\bf{X}} {\bf{X}}^T$, where the $i$-th column of ${\bf{X}}$ is ${\bf{x}}_i - \frac{1}{m}\sum\limits_{j=1}^{m}{\bf{x}}_j$. Then, perform \textit{eigenvalue decomposition} (EVD) to the covariance matrix ${\bf{A}}$ as follows:
\begin{align}
\label{eq_ED}
\begin{split}
{\bf{A}} \bm{v}= \lambda \bm{v} \\
{\bf{X}} {\bf{X}}^T \bm{v}= \lambda \bm{v}
\end{split}
\end{align}
where $\lambda$ is the eigenvalue and $\bm{v}$ is the eigenvector. All eigenvalues make up a $n \times n$ diagonal matrix ${\bf{\Lambda}} = diag(\lambda_1, \lambda_2, \cdots, \lambda_n)$ and all eigenvectors form a $n \times n$ matrix ${\bf{V}} = [\bm{v}_1, \bm{v}_2, \cdots, \bm{v}_n]$.
In order to reduce the dimensionality of the large dataset, the client selects the first 
$d ( d \ll n )$
largest eigenvalues and their corresponding eigenvectors to construct the $n \times d$ projection matrix ${\bf{V}}^L = [\bm{v}^L_1, \bm{v}^L_2, \cdots, \bm{v}^L_d]$, where $\bm{v}^L_i$ is the eigenvector corresponding to the $i$-th largest eigenvalue. Consequently, the $n \times m$ matrix ${\bf{D}}$ is projected to a $d \times m$ truncated matrix $({\bf{V}}^L)^T {\bf{X}}$.

From the PCA process we can see, the core procedure is solving the eigenvalue and eigenvector of the original data's covariance matrix ${\bf{A}}$, which consumes the most computing resources. So for a resource-limited client, the EVD is supposed to be outsourced to the cloud server. 
Moreover, note that computing the original data's covariance matrix ${\bf{A}} = {\bf{X}} {\bf{X}}^T$ is also a time-consuming task. This MMC task can be outsourced as mentioned in Section \ref{sec:4} after the client works out ${\bf{X}}$. And then, we give out the following EVD phases of the outsourcing PCA protocol based on our proposed encryption scheme.

\subsubsection{Secret key generation}
Given the security parameter $\mathcal{K}$, the client selects two random numbers $(\alpha, s)$ and constructs an invertible matrix ${\bf{M}} = {\bf{P}} + {\bf{H}}$ based on our proposed encryption scheme, where ${\bf{P}}(i,j)=p_i \delta_{\pi(i),j}$ and ${\bf{H}} = \left[h_1, \cdots ,h_n \right]^T \left[1, \cdots, 1 \right]$. 

\subsubsection{EVD encryption}
The client encrypts the covariance matrix ${\bf{A}}$ as follows:
\begin{align}
\label{eq_ED_Enc}
\begin{split}
{\bf{A}}' & = \alpha {\bf{A}} + s {\bf{I}}\\
{\bf{B}} & = {\bf{M}}{\bf{A}}'{\bf{M}}^{-1}
\end{split}
\end{align}
After the encryption, the client sends the encrypted EVD task ${\bf{\Phi}}_K({\bf{B}})$ to the cloud server.

\subsubsection{EVD in the cloud} 
On receiving the outsourced EVD task ${\bf{\Phi}}_K({\bf{B}})$, the cloud server performs EVD to the encrypted covariance matrix ${\bf{B}}$ as follows:
\begin{align}
\label{eq_ED_cloud}
{\bf{B}} \bm{v}'= \lambda' \bm{v}'
\end{align}
and then returns the encrypted eigenvalues ${\bf{\Lambda}}' = diag(\lambda'_1, \lambda'_2, \cdots, \lambda'_n)$ and eigenvectors ${\bf{V}}' = [\bm{v}'_1, \bm{v}'_2, \cdots, \bm{v}'_n]$ to the client.

\subsubsection{Result verification}
The equation (\ref{eq_ED_cloud}) can also be written as ${\bf{B}} {\bf{V}}' = {\bf{V}}' {\bf{\Lambda}}'$, and further ${\bf{B}} {\bf{V}}' - {\bf{V}}' {\bf{\Lambda}}' = 0$. This equation will hold only if the cloud server solves ${\bf{\Phi}}_K({\bf{B}})$ honestly. Moreover for the efficiency of verification, the client adopts Monte Carlo verification algorithm \cite{motwani1995randomized} to check whether this equation holds as follows: 
\begin{align}
({\bf{r}}{\bf{B}}) {\bf{V}}' - ({\bf{r}} {\bf{V}}') {\bf{\Lambda}}' = 0
\end{align}
where ${\bf{r}}$ is a random $1 \times n$ dimension $0/1$ vector selected by the client. For a more robust verification, the client can repeat this process in a loop for $l (l \ll n)$ times.

\subsubsection{EVD decryption}
As shown in the equation (\ref{eq_ED_cloud}), it can be transformed as follows:
\begin{align}
\label{eq_ED_dec}
\begin{split}
{\bf{M}}(\alpha {\bf{A}} + s {\bf{I}}){\bf{M}}^{-1} \bm{v}' &= \lambda' \bm{v}' \\
(\alpha {\bf{A}} + s {\bf{I}}){\bf{M}}^{-1} \bm{v}' &= \lambda' {\bf{M}}^{-1} \bm{v}' \\
{\bf{A}} {\bf{M}}^{-1} \bm{v}' + \frac{s}{\alpha} {\bf{M}}^{-1} \bm{v}' &= \frac{\lambda'}{\alpha} {\bf{M}}^{-1} \bm{v}' \\
{\bf{A}} {\bf{M}}^{-1} \bm{v}' &= \frac{\lambda'-s}{\alpha} {\bf{M}}^{-1} \bm{v}'
\end{split}
\end{align}
Therefore according to the equation (\ref{eq_ED}) and (\ref{eq_ED_dec}), the encrypted result $\lambda'$ and $\bm{v}'$ can be decrypted as follows:
\begin{align}
\begin{split}
\lambda &= \frac{\lambda'-s}{\alpha} \\
\bm{v} &= {\bf{M}}^{-1} \bm{v}'
\end{split}
\end{align}

Furthermore, we discuss the rationality of the outsourcing PCA protocol. As shown in the equation (\ref{eq_ED_cloud}), we perform EVD to the encrypted covariance matrix ${\bf{B}}$, while only diagonalizable matrices can be factorized in this way. So we should prove that ${\bf{B}}$ is diagonalizable.

\noindent \textbf{Definition 2.} A square matrix ${\bf{S}}$ is called \textit{diagonalizable} if there exists an invertible matrix ${\bf{T}}$ such that ${\bf{T}}^{-1}{\bf{S}}{\bf{T}}$ is a diagonal matrix.

Since ${\bf{A}} = {\bf{X}} {\bf{X}}^T, {\bf{A}}^T = ({\bf{X}} {\bf{X}}^T)^T =  {\bf{X}} {\bf{X}}^T = {\bf{A}}$, the covariance matrix ${\bf{A}}$ is a real symmetric matrix. ${\bf{A}}'$ is also a real symmetric matrix as all elements of ${\bf{A}}$ are multiplied by $\alpha$ and then only diagonal elements plus a random number $s$. For the reason that the real symmetric matrix is always diagonalizable \cite{zhang2017matrix}, there exists an invertible matrix ${\bf{T}}$ such that ${\bf{T}}^{-1}{\bf{A}}'{\bf{T}}$ is a diagonal matrix. So for the ${\bf{B}} = {\bf{M}}{\bf{A}}'{\bf{M}}^{-1}$, there also exists an invertible matrix ${\bf{M}}{\bf{T}}$ able to transform ${\bf{B}}$ to a diagonal matrix as follows:
\begin{align}
\begin{split}
({\bf{M}}{\bf{T}})^{-1} {\bf{B}} ({\bf{M}}{\bf{T}}) & = {\bf{T}}^{-1}{\bf{M}}^{-1} {\bf{M}}{\bf{A}}'{\bf{M}}^{-1} {\bf{M}}{\bf{T}}\\
& = {\bf{T}}^{-1}{\bf{A}}'{\bf{T}}
\end{split}
\end{align}
Therefore, the encrypted covariance matrix ${\bf{B}}$ is diagonalizable and the proposed outsourcing PCA protocol is rational.

\section{Theoretical Analysis}\label{sec:6}

After presenting our novel encryption scheme based on the subtly designed invertible matrix, we will theoretically analyze whether the design goals are achieved. The correctness has been proved above when we describe the encryption scheme, and moreover we give two examples of applications based on our encryption scheme to show the more general applicability of the proposed encryption scheme. For brevity, we only need to discuss the privacy security, verifiability and efficiency of our proposed scheme.

\subsection{Privacy security analysis}

Throughout the typical outsourcing model, the cloud server only receives encrypted task ${\bf{\Phi}}_K$ from the client and works out the encrypted result ${\bf{\Psi}}_K$ of it. Each secret key $K$ generated by the client will be used to encrypt only one outsourcing task. We analyze the privacy security of our proposed encryption scheme (\textit{i.e.} MMC) in advance.

For the input data, we consider the case given the encrypted matrix ${\bf{X}}' = {\bf{M}}_1{\bf{X}}{\bf{M}}_2^{-1}$, the cloud server attempts to recover the original matrix ${\bf{X}}$. As the computation process of encrypting ${\bf{X}}$ shown in the equation (\ref{X_enc}), we first compute following equations:
\begin{align}
\begin{split}
{\bf{H}}_2 {\bf{P}}^{-1}_2 (i,j) & = \frac{h ^\beta _i}{\beta_j}\\
{\bf{H}}_1 {\bf{X}} {\bf{P}}^{-1}_2 (i,j) & = \frac{h ^\alpha _i}{\beta_j} \sum\limits_{k = 1}^m x_{k,\pi_2(j)}
\end{split}
\end{align}
Consequently, the ciphertext ${\bf{X}}'$ can be expressed as 
\begin{align}
\label{X_ij_enc}
\begin{split}
{\bf{X}}'(i,j) = \frac{\alpha_i}{\beta_j} x_{\pi_1(i),\pi_2(j)} - 
\frac{\alpha_i}{\beta_j(1 + \sum\limits_{k = 1}^n \frac{h^\beta_k}{\beta_k})} \sum\limits_{k = 1}^n \frac{h^\beta_k x_{\pi_1(i),\pi_2(k)}}{\beta_k}\\
\quad + \frac{h ^\alpha _i}{\beta_j} \sum\limits_{k = 1}^m x_{k,\pi_2(j)} - 
\frac{h^\alpha_i}{\beta_j (1 + \sum\limits_{k = 1}^n \frac{h^\beta_k}{\beta_k})} \sum\limits_{l = 1}^n \sum\limits_{k = 1}^m \frac{h^\beta _l}{\beta_l} x_{k,\pi_2 (l)}
\end{split}
\end{align}
In general, except that permutation functions $\pi_1$ and $\pi_2$ perturbs the original data, the cloud server obtains $mn$ nonlinear equations (\ref{X_ij_enc}) consisting of $mn+2m+2n$ unknown variables, resulting in that the cloud server can not work out the original data ${\bf{X}}$ from these equations. Moreover, the ciphertext ${\bf{X}}'$ consists of not only the multiplicative ingredient $\frac{\alpha_i}{\beta_j} x_{\pi_1(i),\pi_2(j)}$ but also other additive ingredients, so the attacker can not know the statistic information of zero elements and exploit it to distinguish the ciphertexts. 

We give a more formal proof as follows:

\noindent \textbf{Proof.} Given two matrices ${\bf{T}} = (t_{ij})$ and ${\bf{T}}' = (t'_{ij})$, the client will generate two sets of secret key $({\bf{M_1}},{\bf{M_2}})$ and $({\bf{M_1}}',{\bf{M_2}}')$ to encrypt them. The ciphertexts ${\bf{E}} = {\bf{M}}_1 {\bf{T}} {\bf{M}}_2^{-1} = (e_{ij})$ and ${\bf{E}}' = {\bf{M}}'_1 {\bf{T}}' {\bf{M}}'^{-1}_2 = (e'_{ij})$ can be represented as
\begin{align*}
\begin{split}
e_{ij} = \frac{\alpha_i}{\beta_j} t_{\pi_1(i),\pi_2(j)} - 
\frac{\alpha_i}{\beta_j(1 + \sum\limits_{k = 1}^n \frac{h^\beta_k}{\beta_k})} \sum\limits_{k = 1}^n \frac{h^\beta_k t_{\pi_1(i),\pi_2(k)}}{\beta_k}\\
\quad + \frac{h ^\alpha _i}{\beta_j} \sum\limits_{k = 1}^m t_{k,\pi_2(j)} - 
\frac{h^\alpha_i}{\beta_j (1 + \sum\limits_{k = 1}^n \frac{h^\beta_k}{\beta_k})} \sum\limits_{l = 1}^n \sum\limits_{k = 1}^m \frac{h^\beta _l}{\beta_l} t_{k,\pi_2 (l)}
\end{split}
\end{align*}
and
\begin{align*}
\begin{split}
e'_{ij} = \frac{\alpha'_i}{\beta'_j} t'_{\pi'_1(i),\pi'_2(j)} - 
\frac{\alpha'_i}{\beta'_j(1 + \sum\limits_{k = 1}^n \frac{h^{\beta'}_k}{\beta'_k})} \sum\limits_{k = 1}^n \frac{h^{\beta'}_k t'_{\pi'_1(i),\pi'_2(k)}}{\beta'_k}\\
\quad + \frac{h ^{\alpha'} _i}{\beta'_j} \sum\limits_{k = 1}^m t'_{k,\pi'_2(j)} - 
\frac{h^{\alpha'}_i}{\beta'_j (1 + \sum\limits_{k = 1}^n \frac{h^{\beta'}_k}{\beta'_k})} \sum\limits_{l = 1}^n \sum\limits_{k = 1}^m \frac{h^{\beta'} _l}{\beta'_l} t'_{k,\pi'_2 (l)}
\end{split}
\end{align*}
Note that the numerical values and permutations of the secret key are randomly generated by the client, and there are additive perturbations hiding zero elements.
So the statistic information of zero elements is protected and the attacker can't rely on it to distinguish the ciphertexts.
In other words, our proposed encryption scheme satisfies IND-ZEA and the inherent weakness of the random-permutation-based encryption scheme has been remedied. $\hfill \blacksquare$

Therefore, the attacker could only try the brute-force attack. The security parameter $\mathcal{K}$ given by the client tends to represent the length of the secret key's element, \textit{i.e.} the random number key space $|\mathcal{P}| = 2^\mathcal{K}$ and $|\mathcal{H}| = 2^\mathcal{K}$. Additionally, with permutation functions, the probability of the attacker working out the key secret by brute-force attack is $\frac{1}{2^{2(m+n)\mathcal{K}}m!n!}$, which is negligible when the scale $m$ and $n$ of the original data ${\bf{X}}$ is large. Further, considering that each secret key $K$ generated by the client will be used to encrypt only one outsourcing task, the cloud cannot recover ${\bf{X}}$ from ${\bf{X}}'$. Likewise, the attacker also can not recover the original data ${\bf{Y}}$ from ${\bf{Y}}'$. Thus the input privacy is protected.

For the output data ${\bf{Z}}$, the cloud can only figure out the encrypted result ${\bf{Z}}' = {\bf{M}}_1 {\bf{Z}} {\bf{M}}_3^{-1}$ by solving the encrypted task ${\bf{\Phi}}_K$. Based on the equation (\ref{X_ij_enc}), we can analyze the protection of output privacy in the same way with that of the input privacy security. Therefore, both the input and the output privacy are protected by the proposed encryption scheme. 

Later, we analyze the privacy security of the other two applications based on the proposed encryption scheme.

\subsubsection{LR based on our encryption scheme} For the input data, $\bm{\mathcal{X}}' = {\bf{A}} \bm{\mathcal{X}} {\bf{M}}$ and $\bm{\mathcal{Y}}' = {\bf{A}} (\bm{\mathcal{Y}} + \bm{\mathcal{X}} {\bm{\mathcal{R}}} )$ are sent to the cloud. 
The ciphertext $\bm{\mathcal{X}}'$ can be represented as 
\begin{align}
\label{LR_ij_enc}
X'_{ij} = \pm k [p_{\pi_3^{-1}(j)} X_{i,\pi_3^{-1}(j)} + \sum\limits_{k = 1}^n h_k X_{ik}]
\end{align}
where $X_{ij}$ and $X'_{ij}$ denote the $i$-th row, $j$-th column element of $\bm{\mathcal{X}}$ and $\bm{\mathcal{X}}'$ respectively.
And the ciphertext $\bm{\mathcal{Y}}'$ can be expressed as 
\begin{align}
\label{LR_i_enc}
Y'_i = \pm k (Y_i + \sum\limits_{k = 1}^n r_k X_{ik})
\end{align}
where $Y_i$ and $Y'_i$ represent the $i$-th row element of $\bm{\mathcal{Y}}$ and $\bm{\mathcal{Y}}'$ respectively.
In general, except that permutation function $\pi$ perturbs the original data, the cloud server can get $mn+m$ nonlinear equations (\ref{LR_ij_enc}) and (\ref{LR_i_enc}) consisting of $mn+m+3n+1$ unknown variables, which implies the cloud server can not recover the original data from these equations. Moreover, the ciphertext consists of both multiplicative and additive transforms, so the number of zero elements is protected and attacks mentioned in \cite{zhou2018efficiently} (\textit{i.e.} greatest common divisor attack and average absolute value attack) are also resisted. Therefore, the attacker could only try the brute-force attack and the probabilities of recovering the original data $\bm{\mathcal{X}}$ and $\bm{\mathcal{Y}}$ are $\frac{1}{2^{(2n+1)\mathcal{K}+m} n!}$ and $\frac{1}{2^{(n+1)\mathcal{K}+m}}$ respectively. For large scale $m$ and $n$, these probabilities are negligible, so the input privacy is protected.

For the encrypted result $\bm{\beta}' = {\bf{M}}^{-1}(\bm{\beta} + {\bm{\mathcal{R}}})$ figured out by the cloud, the regression coefficient $\bm{\beta}$ could be recovered with a negligible probability of $\frac{1}{2^{3n\mathcal{K}} n!}$, which means the output privacy is protected. Thus the input/output privacy security of the outsourcing LR scheme based on the proposed encryption scheme is guaranteed. 


\subsubsection{PCA based on our encryption scheme}
Solving the covariance matrix is an outsourcing MMC task, so the security requirement is guaranteed as mentioned above. Now we should prove the privacy of the outsourcing EVD is protected. 

For the encrypted input data ${\bf{B}} = {\bf{M}}{\bf{A}}'{\bf{M}}^{-1}$ whose element is similar to the equation (\ref{X_ij_enc}), the cloud server could recover the original data with a probability of $\frac{1}{2^{(2n+2)\mathcal{K}} n!}$, which is negligible for large scale $n$. Hence the input privacy is protected.
For the encrypted result $\lambda' = \alpha\lambda + s$ and $\bm{v}' = {\bf{M}} \bm{v}$, probabilities of recovering the eigenvalue $\lambda$ and eigenvector $\bm{v}$ are $\frac{1}{2^{2\mathcal{K}}}$ and $\frac{1}{2^{2n\mathcal{K}} n!}$ respectively. The eigenvector $\bm{v}$ is protected for large scale $n$, and moreover if the security parameter $\mathcal{K}$ is large enough such as $64$-bit, the probability of recovering the eigenvalue $\lambda$ is $\frac{1}{2^{128}}$ which is also negligible. So, the input/output privacy is protected.

\subsection{Verifiability analysis}

For purpose of refusing the false result returned from the cloud server, result verification is necessary for the outsourcing scheme. The proposed encryption scheme applies Monte Carlo verification \cite{motwani1995randomized} and the robust cheating resistance of it has been proved in \cite{lei2013outsourcing,lei2014achieving}, therefore the verifiability requirement is satisfied.

And then we analyze the verifiability of the other two applications based on the proposed encryption scheme.

\subsubsection{LR based on our encryption scheme}
As the equation (\ref{LR_verify}) has shown, the verification equation is equivalent to $\bm{\beta}' = (\bm{\mathcal{X}}'^{T} \bm{\mathcal{X}}')^{-1} \bm{\mathcal{X}}'^{T} \bm{\mathcal{Y}}'$, which is the encrypted task ${\bf{\Phi}}_K$ needs the cloud server to solve. If the encrypted result $\bm{\beta}'$ is false, the verification equation $\bm{\mathcal{X}}'^{T} (\bm{\mathcal{X}}' \bm{\beta}' - \bm{\mathcal{Y}}') \neq 0$ can be detected by the client. So the verification of this LR scheme is robust.

\subsubsection{PCA based on our encryption scheme} The proposed outsourcing PCA scheme also employs Monte Carlo method \cite{motwani1995randomized} to verify the result returned by the cloud server. So this scheme achieves robust cheating resistance.

\subsection{Computation complexity analysis}

As another important design goal, efficiency determines whether the scheme is practical. In this subsection, we analyze the efficiency of the proposed encryption theoretically. Before analyzing the computation complexity of the proposed scheme, we first discuss the computation complexity of multiplying the secret key ${\bf{M}}$ (\textit{i.e.} the subtly designed invertible matrix) by the optimized matrix-chain multiplication method. 

\noindent \textbf{Discussion.} For the sake of brevity, we assume the matrix multiplication condition is always satisfied in the following discussion. Given a general $m \times n$ matrix ${\bf{T}}$ whose 
element is denoted as $t_{ij}$, we use ${\bf{M}}$ to transform it. The subtly designed invertible matrix ${\bf{M}}$ consists of ${\bf{P}}$ and ${\bf{H}}$. Multiplying ${\bf{T}}$ with ${\bf{P}}$ (${\bf{P}}^{-1}$), we can obtain following equations: 
\begin{align}
\label{P_mul}
\begin{split}
{\bf{P}} {\bf{T}} (i,j) & = p_i t_{\pi(i),j} \\
{\bf{T}} {\bf{P}} (i,j) & = p_{\pi^{-1}(j)} t_{i,\pi^{-1}(j)} \\
{\bf{P}}^{-1} {\bf{T}} (i,j) & = \frac{1}{p_{\pi^{-1}(i)}} t_{\pi^{-1}(i),j} \\
{\bf{T}} {\bf{P}}^{-1} (i,j) & = \frac{1}{p_j} t_{i,\pi(j)}
\end{split}
\end{align}
From these equations, we can see the multiplication with a random-permutation-based matrix ${\bf{P}}$ (${\bf{P}}^{-1}$) requires $O(mn)$ computation time. Moreover multiplying ${\bf{T}}$ with ${\bf{H}}$ can be expressed as follows: 
\begin{align}
\label{H_mul}
\begin{split}
{\bf{H}} {\bf{T}} & = [h_1, \cdots, h_m]^T [1, \cdots, 1] \times {\bf{T}} \\
{\bf{T}} {\bf{H}} & = {\bf{T}} \times [h_1, \cdots, h_n]^T [1, \cdots, 1] 
\end{split}
\end{align}
As shown in the equation (\ref{H_mul}), both ${\bf{H}} {\bf{T}}$ and ${\bf{T}} {\bf{H}}$ consist of a matrix-vector and a vector-vector multiplication, costing $O(mn)$ time.
Then the computation complexity of transformation based on the secret key ${\bf{M}}$ (${\bf{M}}^{-1}$) is discussed as follows: 

\noindent \textbf{Left multiply} ${\bf{M}}$: ${\bf{M}} {\bf{T}} = ({\bf{P}} {\bf{T}}) + ({\bf{H}} {\bf{T}})$ consists of one ${\bf{P}}$ multiplication, one ${\bf{H}}$ multiplication and one matrix-matrix addition. So left ${\bf{M}}$ multiplication needs $O(mn)$ time. 

\noindent \textbf{Right multiply} ${\bf{M}}$: ${\bf{T}} {\bf{M}} = ({\bf{T}} {\bf{P}}) + ({\bf{T}} {\bf{H}})$ contains the same operations as left ${\bf{M}}$ multiplication. Thus right ${\bf{M}}$ multiplication also requires $O(mn)$ computation cost. 

\noindent \textbf{Left multiply} ${\bf{M}}^{-1}$: ${\bf{M}}^{-1} {\bf{T}}$ can be represented as $({\bf{P}}^{-1} {\bf{T}}) - \frac{{\bf{P}}^{-1} [{\bf{H}} ({\bf{P}}^{-1} {\bf{T}})]}{1+tr({\bf{H}} {\bf{P}}^{-1})} $, where $tr({\bf{H}} {\bf{P}}^{-1}) = \sum\limits_{k = 1}^m \frac{h_k}{p_k}$ costs $O(m)$ time. Besides working out the trace $tr({\bf{H}} {\bf{P}}^{-1})$, there are still three ${\bf{P}}^{-1}$ multiplications, one ${\bf{H}}$ multiplication and one matrix-matrix subtraction needing to compute. Therefore, the total cost of the left ${\bf{M}}^{-1}$ multiplication is $O(mn)$.

\noindent \textbf{Right multiply} ${\bf{M}}^{-1}$: ${\bf{T}} {\bf{M}}^{-1}$ can be expressed as $({\bf{T}}{\bf{P}}^{-1}) - \frac{[({\bf{T}} {\bf{P}}^{-1}) {\bf{H}}] {\bf{P}}^{-1}}{1+tr({\bf{H}} {\bf{P}}^{-1})} $, which contains the similar operations as ${\bf{M}}^{-1} {\bf{T}}$ except $tr({\bf{H}} {\bf{P}}^{-1}) = \sum\limits_{k = 1}^n \frac{h_k}{p_k}$ costing $O(n)$ time. So, the right ${\bf{M}}^{-1}$ multiplication takes $O(mn)$ time.

\noindent Summarizing the discussion, we can conclude that the multiplication with the secret key ${\bf{M}}$ (${\bf{M}}^{-1}$)
requires $O(mn)$ computation time. $\hfill \blacksquare$

Here we begin to analyze the computation complexity of the proposed encryption scheme formally. 
For the original MMC task ${\bf{Z}} = {\bf{X}} {\bf{Y}}$, the multiplication between $m \times n$ matrix ${\bf{X}}$ and $n \times s$ matrix ${\bf{Y}}$ needs $O(mns)$ computation cost, which generally leads to a huge computation beyond the resource-limited client’s computation power. So the client will resort to cloud computing to complete this task.

To protect the privacy of the original data, ${\bf{X}}$ and ${\bf{Y}}$ have to be encrypted by the client. The client first generates the secret key ${\bf{M}}_1$, ${\bf{M}}_2$ and ${\bf{M}}_3$ by invoking the Algorithm \ref{alg_our_KG}. So the secret key generation costs $O(m+n+s)$ computation time.

Having the secret key, the client has to encrypt the original data ${\bf{X}}$ and ${\bf{Y}}$. ${\bf{X}}$ is transformed based on two ${\bf{M}}$ (${\bf{M}}^{-1}$) multiplications, \textit{i.e.} ${\bf{X}}' = {\bf{M}}_1 {\bf{X}} {\bf{M}}_2^{-1}$. So it takes the client $O(mn)$ to encrypt ${\bf{X}}$. For the reason that encrypting ${\bf{Y}}$ is similar to the ${\bf{X}}$ encryption process, ${\bf{Y}}' = {\bf{M}}_2{\bf{Y}}{\bf{M}}_3^{-1}$ will cost the client $O(ns)$ time. Therefore, the total computation complexity of MMC encryption is $O(mn+ns)$.

After encryption, the encrypted data will be submitted to the cloud server. Once receiving the encrypted data, the cloud server solves the encrypted MMC task by multiplying ${\bf{X}}'$ and ${\bf{Y}}'$ to get the encrypted result ${\bf{Z}}'$. The computation cost of ${\bf{Z}}' = {\bf{X}}' {\bf{Y}}'$ is $O(mns)$, same as that of solving the original MMC task without other additional overhead.

Then, the result ciphertext ${\bf{Z}}'$ will be sent to the client and decrypted using the secret key. The MMC decryption ${\bf{Z}} = {\bf{M}}_1^{-1} {\bf{Z}}' {\bf{M}}_3$ consists of two ${\bf{M}}$ (${\bf{M}}^{-1}$) multiplications. So this process costs $O(ms)$ time.

In order to resist a false result, the client executes the Monte Carlo verification algorithm ${\bf{R}} = {\bf{X}} \times ({\bf{Yr}}) - {\bf{Zr}}$ in a loop for $l$
times. Since $l \ll m,n,s$ and each verification contains three matrix-vector multiplications, the overall cost of the result verification process is $O(mn+ns+ms)$.

\begin{table}[ht]
\centering
\caption{Comparison of the computation complexity between our proposed scheme and Lei et al.'s scheme \cite{lei2014achieving}}
\label{comp_MMC}
\begin{threeparttable}
\begin{tabular}{c|c|c|c}
\Xhline{1pt}
\textbf{Party} & \textbf{Phase} & \thead{\textbf{Lei et al.’s} \\ \textbf{scheme \cite{lei2014achieving}}} & \thead{\textbf{Our proposed} \\ \textbf{scheme}} \\
\Xhline{0.75pt}
  \multirow{9}*{\textbf{Client}}  
    & \thead{secret key \\ generation} &    $O(m+n+s)$    & $O(m+n+s)$ \\
    \cline{2-4}          
    & \thead{MMC \\ encryption}   &   $O(mn+ns)$    &  $O(mn+ns)$ \\   
    \cline{2-4} 
    & \thead{MMC \\ decryption} &   $O(ms)$    &  $O(ms)$ \\
    \cline{2-4} 
    & \thead{result \\ verification}  &   $O(mn+ns+ms)$    &  $O(mn+ns+ms)$ \\      
\Xhline{0.75pt}         
  \thead{\textbf{Cloud} \\ \textbf{Server}} 
    & \thead{MMC in \\ the cloud}   &   $O(mns)$    & $O(mns)$ \\
\Xhline{0.75pt}  
  \multicolumn{2}{c|}{\textbf{Privacy Security}} & $\times$ & $\surd$ \\ 
  \cline{1-4}
  \multicolumn{2}{c|}{\textbf{Applicability}} & $\surd$ & $\surd$ \\
\Xhline{1pt} 
\end{tabular}
\begin{tablenotes}
  \footnotesize
  \item[*] The computation complexity of original MMC is $O(mns)$.
  \item[*] The $\surd$ of \textbf{Privacy Security} means a higher security.
  \item[*] The $\surd$ of \textbf{Applicability} means a wider applicability.
\end{tablenotes}
\end{threeparttable}
\end{table}

For clarity, we list the comparison between our proposed scheme and Lei et al.'s scheme \cite{lei2014achieving} in the Table \ref{comp_MMC}. The overall computation complexity is $O(mn+ns+ms)$ for the client and $O(mns)$ for the cloud. In terms of efficiency, the proposed scheme is practical on account of the gap between $O(mn+ns+ms)$ and $O(mns)$. As long as $(m,n,s)$ are sufficiently large, the proposed scheme is able to allow the client to outsource MMC task to the cloud server and gain substantial computation savings. This claim will be further validated by experiments in the next section.


In summary, besides the enhanced privacy security provided by our proposed scheme, the efficiency of this scheme is comparable to Lei et al.'s scheme \cite{lei2014achieving} and makes the computation overload of the client reduced substantially by outsourcing.

Further, we analyze the computation complexity of the other two applications based on the proposed encryption scheme, and compare the performance of these applications with that of state-of-the-art schemes based on matrix transformation.

\subsubsection{LR based on our encryption scheme}
Before submitting LR task to cloud server, the client has to generate secret key and encrypt the original data $\bm{\mathcal{X}}$ and $\bm{\mathcal{Y}}$. The secret key consists of $n+1$ random numbers $(k, {\bm{\mathcal{R}}})$ and a $n \times n$ subtly designed invertible matrix ${\bf{M}}$ generated by Algorithm \ref{alg_our_KG}. So the secret key generation costs $O(n)$ time. The encryption process $\bm{\mathcal{X}}' = {\bf{A}} \bm{\mathcal{X}} {\bf{M}}$ contains one diagonal matrix multiplication and one ${\bf{M}}$ multiplication, and $\bm{\mathcal{Y}}' = {\bf{A}} (\bm{\mathcal{Y}} + \bm{\mathcal{X}} {\bm{\mathcal{R}}})$ consists of one matrix-verctor multiplication, one vector-vector addition and one diagonal matrix multiplication. Hence, the overall LR encryption phase takes $O(mn)$ computation time.

On receipt of the encrypted data $\bm{\mathcal{X}}'$ and $\bm{\mathcal{Y}}'$, the cloud server computes $\bm{\beta}' = (\bm{\mathcal{X}}'^{T} \bm{\mathcal{X}}')^{-1} \bm{\mathcal{X}}'^{T} \bm{\mathcal{Y}}'$, of which the computation complexity is same as that of the original LR task $\bm{\beta} = (\bm{\mathcal{X}}^T \bm{\mathcal{X}})^{-1} \bm{\mathcal{X}}^T \bm{\mathcal{Y}}$ containing one matrix-matrix multiplication, one matrix inversion and two matrix-vector multiplications. So, the LR in the cloud phase costs $O(mn^2 + n^3)$ time.

Then the encrypted result $\bm{\beta}'$ returned from the cloud server to the client will be verified by checking whether $\bm{\mathcal{X}}'^{T} (\bm{\mathcal{X}}' \bm{\beta}' - \bm{\mathcal{Y}}')$ equals to $0$. This process consists of two matrix-vector multiplications and one vector-vector subtraction. So the time cost of result verification is $O(mn)$.

If the verification succeeds, the client will decrypt the encrypted regression coefficient $\bm{\beta} = {\bf{M}} \bm{\beta}' - {\bm{\mathcal{R}}}$ and work out the intercept term $\beta_0 = \bar{y} - \bar{\bf{x}}\bm{\beta}$. This process consists of one matrix-vector multiplication, one vector-vector multiplication, one vector-vector subtraction and one scalar subtraction. So the LR decryption takes $O(n^2)$ time.

\begin{table}[ht]
\centering
\caption{Comparison between the proposed LR scheme and Zhou et al.'s scheme \cite{zhou2018efficiently}}
\label{comp_LR}
\begin{threeparttable}
\begin{tabular}{c|c|c|c}
\Xhline{1pt}
\textbf{Party} & \textbf{Phase} & \thead{\textbf{Zhou et al.’s} \\ \textbf{scheme \cite{zhou2018efficiently}}} & \thead{\textbf{Our proposed} \\ \textbf{scheme}} \\
\Xhline{0.75pt}
  \multirow{6}*{\textbf{Client}}  
    & \thead{secret key  generation} &    $O(n)$    & $O(n)$ \\
    \cline{2-4}          
    & \thead{LR  encryption} &   $O(mn)$    &  $O(mn)$ \\   
    \cline{2-4} 
    & \thead{result  verification} &   $O(mn)$    &  $O(mn)$ \\
    \cline{2-4} 
    & \thead{LR  decryption} &  $O(n^2)$    &  $O(n^2)$ \\ 
\Xhline{0.75pt}         
  \thead{\textbf{Cloud} \\ \textbf{Server}} 
    & \thead{LR in  the cloud} &   $O(mn^2 + n^3)$    & $O(mn^2 + n^3)$ \\  
\Xhline{0.75pt}  
  \multicolumn{2}{c|}{\textbf{Privacy Security}} & $\surd$ & $\surd$ \\ 
  \cline{1-4}
  \multicolumn{2}{c|}{\textbf{Applicability}} & $\times$ & $\surd$ \\
\Xhline{1pt} 
\end{tabular}
\begin{tablenotes}
  \footnotesize
  \item[*] The computation complexity of original LR is $O(mn^2 + n^3)$.
  \item[*] The $\surd$ of \textbf{Privacy Security} means a higher security.
  \item[*] The $\surd$ of \textbf{Applicability} means a wider applicability.
\end{tablenotes}
\end{threeparttable}
\end{table}

We compare the proposed outsourcing LR scheme to the state-of-the-art scheme based on matrix transformation \cite{zhou2018efficiently} and list the comparison in Tabel \ref{comp_LR} for clarity. This outsourcing LR protocol based on our proposed encryption scheme not only protects the privacy of original data, but also reduces the overhead of client from $O(mn^2+n^3)$ to $O(mn+n^2)$, making it possible for the resource-constrained client to solve a large-scale LR task. Moreover, the performance of Zhou et al.s scheme \cite{zhou2018efficiently} is same as this outsourcing LR protocol, but their encryption method may be not suitable for a situation where the decryption needs the inverse of the secret key as mentioned in Section \ref{sec:1}.


\subsubsection{PCA based on our encryption scheme}
As mentioned above, the original data’s covariance matrix ${\bf{A}} = {\bf{X}} {\bf{X}}^T$ can be solved by the MMC scheme. For brevity, we only analyze the computation complexity of outsourcing EVD phases. 

To generate the secret key, the client selects two random numbers $(\alpha, s)$ and constructs an invertible matrix ${\bf{M}}$ based on the proposed encryption scheme. So the complexity of secret key generation is $O(n)$.

The encryption process ${\bf{A}}' = \alpha {\bf{A}} + s {\bf{I}}$ and ${\bf{B}} = {\bf{M}}{\bf{A}}'{\bf{M}}^{-1}$ consist of one scalar-matrix multiplication, one diagonal matrix addition and two ${\bf{M}}$ (${\bf{M}}^{-1}$) multiplications. So the EVD encryption needs $O(n^2)$ computation time.

After receiving the encrypted data ${\bf{B}}$, the cloud server executes EVD of it without any other additional overhead compared to the original EVD. Thus, EVD in the cloud will cost $O(n^3)$ time.

For the verification, the client adopts the Monte Carlo verification algorithm $({\bf{r}}{\bf{B}}) {\bf{V}}' - ({\bf{r}} {\bf{V}}') {\bf{\Lambda}}' = 0$ in a loop for $l$ times. As $l \ll n$ and each verification contains three matrix-vector multiplications and one diagonal matrix multiplication, the overall complexity of the result verification is $O(n^2)$.

After passing verification, the encrypted eigenvalues ${\bf{\Lambda}}' = diag(\lambda'_1, \lambda'_2, \cdots, \lambda'_n)$ and eigenvectors ${\bf{V}}' = [\bm{v}'_1, \bm{v}'_2, \cdots, \bm{v}'_n]$ will be decrypted by computing ${\bf{\Lambda}} = \frac{1}{\alpha} ({\bf{\Lambda}}'-s{\bf{I}})$ and ${\bf{V}} = {\bf{M}}^{-1} {\bf{V}}'$. This process contains one diagonal matrix subtraction and one ${\bf{M}}^{-1}$ multiplication. So the computation cost of EVD decryption is $O(n^2)$.

\begin{table}[ht]
\centering
\caption{Comparison between the proposed EVD method and Zhou et al.'s scheme \cite{zhou2016outsourcing}}
\label{comp_EVD}
\begin{threeparttable}
\begin{tabular}{c|c|c|c}
\Xhline{1pt}
\textbf{Party} & \textbf{Phase} & \thead{\textbf{Zhou et al.’s} \\ \textbf{scheme \cite{zhou2016outsourcing}}} & \thead{\textbf{Our proposed} \\ \textbf{scheme}} \\
\Xhline{0.75pt}
  \multirow{6}*{\textbf{Client}}  
    & \thead{secret key  generation} &  $O(n)$    & $O(n)$ \\
    \cline{2-4}          
    & \thead{EVD  encryption} &  $O(n^2)$    &  $O(n^2)$ \\   
    \cline{2-4} 
    & \thead{result  verification} &    $O(n^2)$    &  $O(n^2)$ \\
    \cline{2-4} 
    & \thead{EVD  decryption} &   $O(n^2)$    &  $O(n^2)$ \\ 
\Xhline{0.75pt}         
  \thead{\textbf{Cloud} \\ \textbf{Server}} 
    & \thead{EVD in  the cloud} &   $O(n^3)$    & $O(n^3)$ \\  
\Xhline{0.75pt}  
  \multicolumn{2}{c|}{\textbf{Privacy Security}} & $\times$ & $\surd$ \\ 
  \cline{1-4}
  \multicolumn{2}{c|}{\textbf{Applicability}} & $\surd$ & $\surd$ \\
\Xhline{1pt} 
\end{tabular}
\begin{tablenotes}
  \footnotesize
  \item[*] The computation complexity of original EVD is $O(n^3)$.
  \item[*] The $\surd$ of \textbf{Privacy Security} means a higher security.
  \item[*] The $\surd$ of \textbf{Applicability} means a wider applicability.
\end{tablenotes}
\end{threeparttable}
\end{table}

In order to compare the proposed outsourcing EVD method with the state-of-the-art method based on matrix transformation \cite{zhou2016outsourcing} more clearly, we list the comparison in Table \ref{comp_EVD}. This outsourcing EVD method based on our proposed encryption scheme not only protects the privacy of original data, but also reduces the computation complexity of client from $O(n^3)$ to $O(n^2)$ so that the resource-constrained client can complete the EVD phases of PCA. Though the efficiency of Zhou et al.s scheme \cite{zhou2016outsourcing} is comparable to this outsourcing EVD, but they employ the random-permutation-based encryption scheme similar to Lei et al's scheme, which leads to the inherent security weakness mentioned in Section \ref{sec:3}.


\noindent \textbf{Remark.} Due to sizes of encrypted data ${\bf{\Phi}}_K$ sent to the cloud and result ${\bf{\Psi}}_K$ returned to the client are same as those of the original data $\bf{\Phi}$ and result $\bf{\Psi}$, the proposed scheme doesn't introduce extra communication overhead.

\section{Performance Evaluation}\label{sec:7}

In the previous section, we analyze the efficiency of the proposed encryption scheme and other applications theoretically. In order to intuitively present the performance of these schemes, we fully implement them to evaluate the practicality numerically in this section. The experimental result indicates that the proposed encryption scheme based on the subtly designed invertible matrix is practical and comparable to the state-of-the-art schemes based on matrix transformation.

\begin{figure*}[ht]
\centering
\subfigure[Trend of client speedup $i_{\rm c}$ with dimension $n$]{
  \includegraphics[width=8cm]{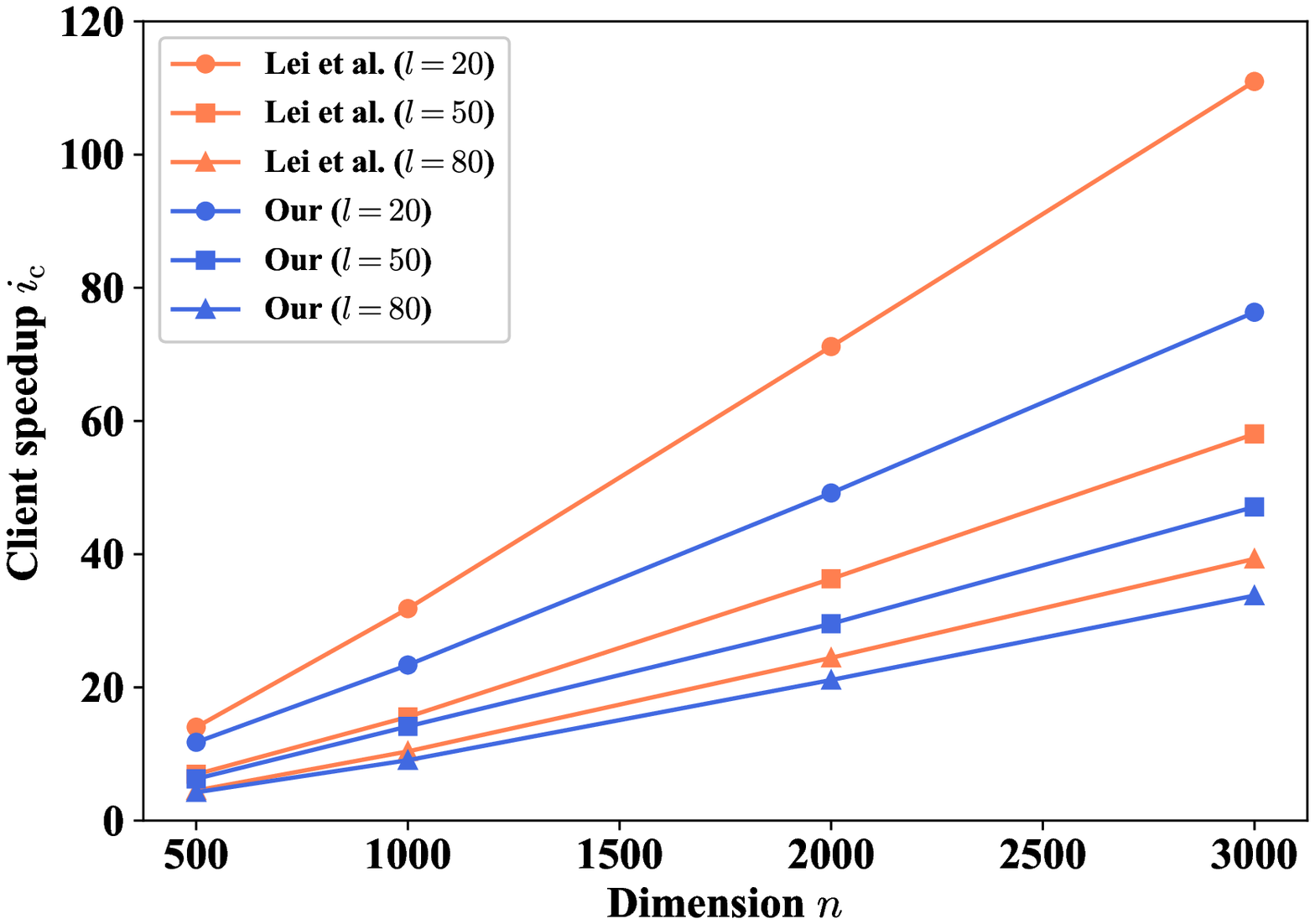}
}
\quad
\subfigure[Trend of relative extra cost $i_{\rm ec}$ with dimension $n$]{
  \includegraphics[width=8cm]{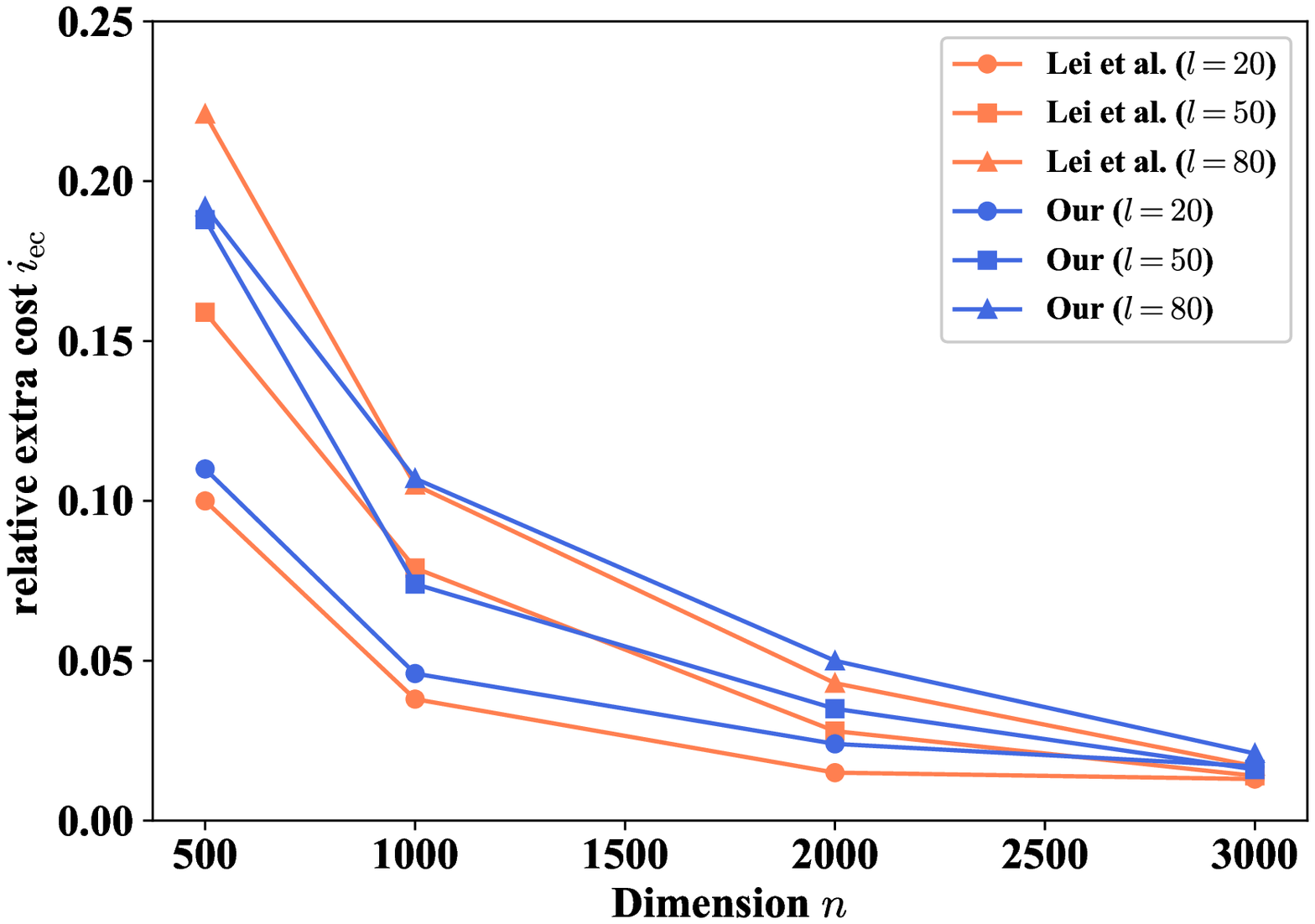}
}
\caption{Trend of client speedup $i_{\rm c}$ and relative extra cost $i_{\rm ec}$ with the scale of data ($m:n:s = 4:5:6$)}
\label{Exp1_Fig}
\end{figure*}

\subsection{Experiment setup}

To evaluate the performance of the proposed encryption scheme and other applications based on it, 
we implement them using C++ as the programming language. All experiments are conducted on a Windows 10 machine with 4-core 3.40GHz Intel i7 CPU and 24GB RAM to simulate the process on client and cloud server. As mentioned in \cite{lei2013outsourcing,lei2014achieving}, if we implement the experiment for both client side and cloud side on the same machine and measure their running time, then the result can reflect the asymmetric amount of computation performed in both sides.
However, if we implement the experiment on two different machines with one in client side and the other in cloud side, then the result may depend on the asymmetric computing speed owned by the two different machines. 
Therefore, one-machine-based experiment is employed. 
Additionally, due to the proposed scheme introduces no extra communication overhead, we don't need to consider the communication latency between the client and the cloud.

Moreover for the sake of clarity, we list the main notations used throughout the experiment in Table \ref{note_exp}.

\begin{table}[ht]
\centering
\caption{Notations and definitions in the experiment}
\label{note_exp}
\begin{tabular}{c l}
  \Xhline{1pt}
  \textbf{Notation} & \textbf{Definition} \\
      \hline
      $t_{\rm o}$ & \tabincell{l}{The time for computing the original task} \bigstrut \\
      $t_{\rm cs}$ & \tabincell{l}{The time for the cloud server to compute the enc- \\rypted task} \bigstrut \\
      $t_{\rm c1}$ & \tabincell{l}{The time for client to generate the secret key and \\ encrypt the original task} \bigstrut \\
      $t_{\rm c2}$ & \tabincell{l}{The time for the client to decrypt and verify the \\ returned result} \bigstrut \\
      $t_{\rm c}$ & \tabincell{l}{$t_{\rm c} = t_{\rm c1} + t_{\rm c2}$} \bigstrut \\
      $i_{\rm c}$ & \tabincell{l}{A performance indicator called client speedup} \bigstrut \\
      $i_{\rm cs}$ & \tabincell{l}{A performance indicator called cloud efficiency} \bigstrut \\
      $i_{\rm ec}$ & \tabincell{l}{A performance indicator called relative extra cost} \bigstrut \\
  \Xhline{1pt}
\end{tabular}
\end{table}

\begin{table*}[ht]
\centering
\caption{Performance evaluation of encryption schemes (Time is in milliseconds, \textit{i.e.} ms)}
\label{Exp_1_table}
\begin{tabular}{c|c|c|c|c|c|c|c|c|c|c|c|c|c|c}
\hline
    \multirow{2}[4]{*}{\textbf{\tabincell{c}{Verification \\ loop}}} & \multirow{2}[4]{*}{$\bm{(m,n,s)$}} &  \textbf{\tabincell{c}{Original \\ MMC}} & \multicolumn{6}{c|}{\textbf{Lei et al.'s scheme \cite{lei2014achieving}}} & \multicolumn{6}{c}{\textbf{Our proposed scheme}} \bigstrut\\
\cline{3-15}          &       & $t_{\rm o}$ & $t_{\rm cs}$ & $t_{\rm c1}$ & $t_{\rm c2}$ & $i_{\rm c}$  & $i_{\rm cs}$   & $i_{\rm ec}$  & $t_{\rm cs}$ & $t_{\rm c1}$ & $t_{\rm c2}$ & $i_{\rm c}$    & $i_{\rm cs}$    & $i_{\rm ec}$ \bigstrut\\
    \hline
    \multirow{4}[8]{*}{$\bm{l=20}$} & (400,500,600) & 1094  & 1125  & 16    & 62    & 14.03 & 0.97  & 0.100  & 1121  & 31    & 62    & 11.76 & 0.98  & 0.110 \bigstrut\\
\cline{2-15}          & (800,1000,1200) & 9485  & 9546  & 63    & 235   & 31.83 & 0.99  & 0.038  & 9516  & 140   & 266   & 23.36 & 1.00  & 0.046 \bigstrut\\
\cline{2-15}          & (1600,2000,2400) & 84563 & 84656 & 281   & 907   & 71.18 & 1.00  & 0.015  & 84859 & 609   & 1110  & 49.19 & 1.00  & 0.024 \bigstrut\\
\cline{2-15}          & (2400,3000,3600) & 301750 & 302859 & 672   & 2047  & 110.98 & 1.00  & 0.013  & 302906 & 1422  & 2531  & 76.33 & 1.00  & 0.017 \bigstrut\\
    \hline
    \multirow{4}[8]{*}{$\bm{l=50}$} & (400,500,600) & 1078  & 1094  & 15    & 140   & 6.95  & 0.99  & 0.159  & 1109  & 31    & 141   & 6.27  & 0.97  & 0.188 \bigstrut\\
\cline{2-15}          & (800,1000,1200) & 9500  & 9640  & 79    & 531   & 15.57 & 0.99  & 0.079  & 9528  & 125   & 547   & 14.14 & 1.00  & 0.074 \bigstrut\\
\cline{2-15}          & (1600,2000,2400) & 84516 & 84594 & 281   & 2047  & 36.30 & 1.00  & 0.028  & 84610 & 609   & 2250  & 29.56 & 1.00  & 0.035 \bigstrut\\
\cline{2-15}          & (2400,3000,3600) & 303859 & 302828 & 672   & 4562  & 58.05 & 1.00  & 0.014  & 302234 & 1406  & 5046  & 47.10 & 1.01  & 0.016 \bigstrut\\
    \hline
    \multirow{4}[8]{*}{$\bm{l=80}$} & (400,500,600) & 1062  & 1062  & 16    & 219   & 4.52  & 1.00  & 0.221  & 1016  & 31    & 219   & 4.25  & 1.05  & 0.192 \bigstrut\\
\cline{2-15}          & (800,1000,1200) & 8937  & 9015  & 63    & 797   & 10.39 & 0.99  & 0.105  & 8906  & 141   & 844   & 9.07  & 1.00  & 0.107 \bigstrut\\
\cline{2-15}          & (1600,2000,2400) & 84421 & 84625 & 297   & 3157  & 24.44 & 1.00  & 0.043  & 84672 & 625   & 3375  & 21.11 & 1.00  & 0.050 \bigstrut\\
\cline{2-15}          & (2400,3000,3600) & 305140 & 302688 & 672   & 7093  & 39.30 & 1.01  & 0.017  & 302484 & 1438  & 7593  & 33.79 & 1.01  & 0.021 \bigstrut\\
\hline
\end{tabular}%
\end{table*}%

\begin{table*}[htbp]
  \centering
  \caption{Performance evaluation of LR schemes (Time is in milliseconds, \textit{i.e.} ms)}
  \label{Exp_2_table}%
    \begin{tabular}{c|c|c|c|c|c|c|c|c|c|c|c|c|c}
    \hline
    \multirow{2}[4]{*}{$\bm{(m,n)}$} & \textbf{Original LR} & \multicolumn{6}{c|}{\textbf{Zhou et al.'s LR scheme \cite{zhou2018efficiently}}}  & \multicolumn{6}{c}{\textbf{LR scheme based on the proposed encryption scheme}} \bigstrut\\
\cline{2-14}          & $t_{\rm o}$ & $t_{\rm cs}$ & $t_{\rm c1}$ & $t_{\rm c2}$ & $i_{\rm c}$ & $i_{\rm cs}$ & $i_{\rm ec}$ & $t_{\rm cs}$ & $t_{\rm c1}$ & $t_{\rm c2}$ & $i_{\rm c}$ & $i_{\rm cs}$ & $i_{\rm ec}$ \bigstrut\\
    \hline
    (600,300) & 3625  & 3655  & 15    & 0     & 241.67 & 0.99  & 0.012 & 3626  & 15    & 0     & 241.67 & 1.00  & 0.004 \bigstrut\\
    \hline
    (600,400) & 7795  & 7813  & 16    & 0     & 487.19 & 1.00  & 0.004 & 7796  & 16    & 0     & 487.19 & 1.00  & 0.002 \bigstrut\\
    \hline
    (600,600) & 23940 & 24078 & 16    & 0     & 1496.25 & 0.99  & 0.006 & 23734 & 15    & 0     & 1596.00 & 1.01  & -0.008 \bigstrut\\
    \hline
    (1000,600) & 27531 & 28389 & 32    & 16    & 573.56 & 0.97  & 0.033 & 27642 & 47    & 0     & 585.77 & 1.00  & 0.006 \bigstrut\\
    \hline
    (2000,600) & 36297 & 36390 & 78    & 16    & 386.14 & 1.00  & 0.005 & 36157 & 93    & 15    & 336.08 & 1.00  & -0.001 \bigstrut\\
    \hline
    (3000,600) & 42312 & 42876 & 125   & 16    & 300.09 & 0.99  & 0.017 & 42624 & 141   & 15    & 271.23 & 0.99  & 0.011 \bigstrut\\
    \hline
    (4000,600) & 50061 & 50281 & 156   & 15    & 292.75 & 1.00  & 0.008 & 49782 & 171   & 16    & 267.71 & 1.01  & -0.002 \bigstrut\\
    \hline
    (5000,600) & 56985 & 57328 & 172   & 31    & 280.71 & 0.99  & 0.010 & 56688 & 234   & 16    & 227.94 & 1.01  & -0.001 \bigstrut\\
    \hline
    (6000,600) & 65608 & 65218 & 219   & 31    & 262.43 & 1.01  & -0.002 & 64656 & 281   & 16    & 220.90 & 1.01  & -0.010 \bigstrut\\
    \hline
    \end{tabular}%
\end{table*}%

\begin{table*}[htbp]
  \centering
  \caption{Performance evaluation of EVD schemes (Time is in milliseconds, \textit{i.e.} ms)}
  \label{Exp_3_table}%
    \begin{tabular}{c|c|c|c|c|c|c|c|c|c|c|c|c|c}
    \hline
    \multirow{2}[4]{*}{$\bm{n}$} & \textbf{Original EVD} & \multicolumn{6}{c|}{\textbf{Zhou et al.'s EVD scheme \cite{zhou2016outsourcing}}} & \multicolumn{6}{c}{\textbf{EVD scheme based on the proposed encryption scheme}} \bigstrut\\
\cline{2-14}          & $t_{\rm o}$ & $t_{\rm cs}$ & $t_{\rm c1}$ & $t_{\rm c2}$ & $i_{\rm c}$ & $i_{\rm cs}$ & $i_{\rm ec}$ & $t_{\rm cs}$ & $t_{\rm c1}$ & $t_{\rm c2}$ & $i_{\rm c}$ & $i_{\rm cs}$ & $i_{\rm ec}$ \bigstrut\\
    \hline
    600   & 145919 & 146486 & 12    & 469   & 303.37 & 1.00  & 0.007 & 161069 & 19    & 472   & 297.19 & 0.91  & 0.107 \bigstrut\\
    \hline
    1200  & 1097992 & 1101799 & 55    & 2599  & 413.71 & 1.00  & 0.006 & 1207110 & 98    & 2603  & 406.51 & 0.91  & 0.102 \bigstrut\\
    \hline
    1800  & 3656864 & 3673483 & 131   & 6506  & 550.98 & 1.00  & 0.006 & 3994437 & 253   & 6653  & 529.52 & 0.92  & 0.094 \bigstrut\\
    \hline
    2400  & 8550123 & 8581153 & 257   & 11795 & 709.44 & 1.00  & 0.005 & 9443053 & 451   & 11909 & 691.76 & 0.91  & 0.106 \bigstrut\\
    \hline
    \end{tabular}%
\end{table*}%

\subsection{The proposed encryption scheme}

The goal of experiment is to present the performance improvement for the client by outsourcing. So, the main performance indicator can be expressed as a ratio of the time needed if the original task is done locally over the time needed by the client if outsourcing is chosen, \textit{i.e.} $i_{\rm c} = \frac{t_{\rm o}}{t_{\rm c}}$ which is referred to as \textit{client speedup}. This value should be a considerable positive number greater than $1$, which implies the computation overload of client is reduced substantially by outsourcing. 
Another metric $i_{\rm cs} = \frac{t_{\rm o}}{t_{\rm cs}}$ called \textit{cloud efficiency} is also considered, \textit{i.e.} the overhead introduced by the encryption scheme to solve the task in the cloud server. This value is expected to be close to $1$, which means the encryption scheme should not increase time cost to solve the encrypted task compared with the original one in the cloud.
Moreover, we use \textit{relative extra cost} $ i_{\rm ec} = \frac{t_{\rm c} + t_{\rm cs} - t_{\rm o}}{t_{\rm o}}$ to measure the total extra cost at both client and cloud server.

To present the performance evaluation clearly, we list results in Table \ref{Exp_1_table} where the runtime is the average value of 20 runs (in  milliseconds, \textit{i.e.} ms). We set the matrix dimensions $m:n:s = 4:5:6$ and conduct experiments with different verification loops $l = 20$, $l=50$ and $l=80$, which correspond to efficiency priority, tradeoff, and cheating resistance priority cases. As shown in Table \ref{Exp_1_table}, the cloud efficiency $i_{cs}$ is close to $1$, indicating the encryption scheme barely increases time cost to solve the encrypted task compared with the original one in the cloud. And $t_{c1} < t_{c2}$, which means handling result decryption and verification takes longer than generating secret key and encrypting the original data. 

To show the trend of client speedup $i_{\rm c}$ and relative extra cost $i_{\rm ec}$ with the scale of data more intuitively, we present it in Fig. \ref{Exp1_Fig}. It is shown in this figure that the client speedup $i_{\rm c}$ increases along with the growth of the scale of data and accordingly the relative extra cost $i_{\rm ec}$ becomes basically lower. Moreover with the verification loop increasing, $i_{\rm c}$ decreases and $i_{\rm ec}$ becomes generally higher. Furthermore compared with Lei et al.'s encryption scheme based on random permutation \cite{lei2014achieving}, performance of our proposed encryption scheme falls but is still comparable to Lei et al.'s scheme. This is reasonable that the proposed scheme involves both multiplicative and additive perturbation to solve the inherent security weakness of Lei et al.'s scheme mentioned in Section \ref{sec:3}. Therefore, besides having the enhanced security, our proposed encryption scheme is also efficient enough in practice.

\subsection{The outsourcing LR scheme}

The experiment result is listed in Table \ref{Exp_2_table}. As shown in this table, the cloud efficiency $i_{\rm cs}$ approximately equals to $1$, meaning the encryption scheme doesn't introduce additional overhead to solve the encrypted LR compared to the original LR in the cloud. 
And $t_{c1} < t_{c2}$, which implies result decryption and verification costs more time than secret key generation and LR encryption. Moreover, the relative extra cost $i_{\rm ec}$ is very small, no more than $2\%$ in general. For the client speedup $i_{\rm c}$, we can see it is related to both $m$ and $n$. When $m$ is set a fixed value (\textit{e.g.} $m=600$), $i_{\rm c}$ increases with the growth of $n$. If the value of $n$ is fixed (\textit{e.g.} $n=600$), $i_{\rm c}$ will be lower with $m$ increasing. Further the outsourcing LR scheme based on our proposed encryption scheme has a comparable performance with Zhou et al.'s scheme.

\subsection{The outsourcing PCA scheme}

As mentioned above, PCA consists of computing the original data’s covariance matrix ${\bf{A}} = {\bf{X}} {\bf{X}}^T$ and conducting EVD to ${\bf{A}}$. The covariance matrix ${\bf{A}}$ can be solved by the outsourcing MMC scheme. To avoid repetition, we only simulate the outsourcing EVD phases. The simulation result is given in Table \ref{Exp_3_table}. From this table, we can see $i_{\rm cs}$ nearly equals to $0.9$ less than $1$. This is because conducting EVD to a real symmetric matrix is more efficient than a general diagonalizable matrix but the encrypted covariance matrix ${\bf{B}} = {\bf{M}}{\bf{A}}'{\bf{M}}^{-1}$ is not a real symmetric matrix. Though the loss of cloud efficiency is acceptable, this problem still deserves further study. Consequently the relative extra cost $i_{\rm ec}$ is generally about $10\%$. 
Compared with the $i_{\rm cs}$ of Zhou et al.'s scheme \cite{zhou2016outsourcing} equaling to $1$, they employed the random permutation matrix and set all random numbers of it to $1$ or $-1$ to keep the random permutation matrix orthogonal so that the encrypted covariance matrix is still symmetric. However, the security of \cite{zhou2016outsourcing} is even inferior to Lei et al.'s work \cite{lei2014achieving}.
Moreover the client speedup $i_{\rm c}$ increases with the growth of the dimension $n$ and meets the practical needs. 

All in all, besides the enhanced security compared to the random-permutation-based encryption scheme, our proposed encryption scheme based on the subtly designed invertible matrix can be widely used in various outsourced tasks and the performance is practical.

\section{Conclusion}\label{sec:8}

In this paper, we have proposed a security-enhanced encryption scheme based on subtly designed invertible matrix for privacy-preserving cloud computing, solving the inherent weakness of the random-permutation-based encryption scheme \textit{i.e.} revealing the statistic information of zero elements in the original data and not satisfying IND-ZEA. Further, the proposed scheme can be applied to not only MMC but also other varieties of outsourced large-scale tasks with comparable performance to the state-of-the-art scheme based on matrix transformation. Moreover, there exists several efficient and robust methods to verify the correctness of the result returned by the cloud server such as Monte Carlo verification algorithm.

\ifCLASSOPTIONcaptionsoff
  \newpage
\fi


\bibliographystyle{IEEEtran}
\bibliography{IEEEabrv,MyPaper20200301}

\begin{IEEEbiography}[{\includegraphics[width=1in,height=1.25in,clip,keepaspectratio]{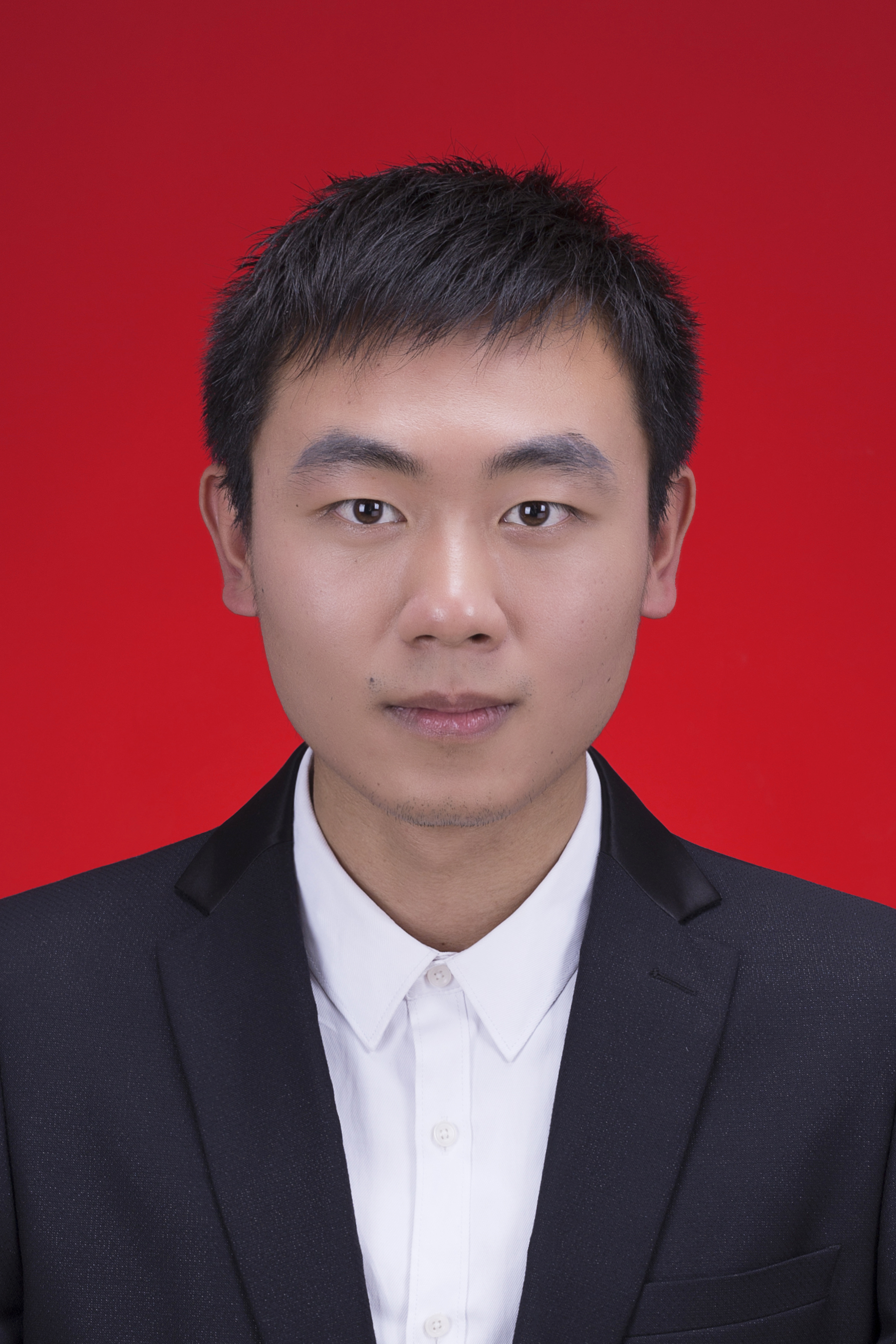}}]{Chun Liu}
received B.E. degree from Zhejiang University, Hangzhou, China, in 2017. He is working towards the Master degree in the State Key Laboratory of Mathematical Engineering and Advanced Computing. His research interests include applied cryptography and big data security.
\end{IEEEbiography}

\begin{IEEEbiography}[{\includegraphics[width=1in,height=1.25in,clip,keepaspectratio]{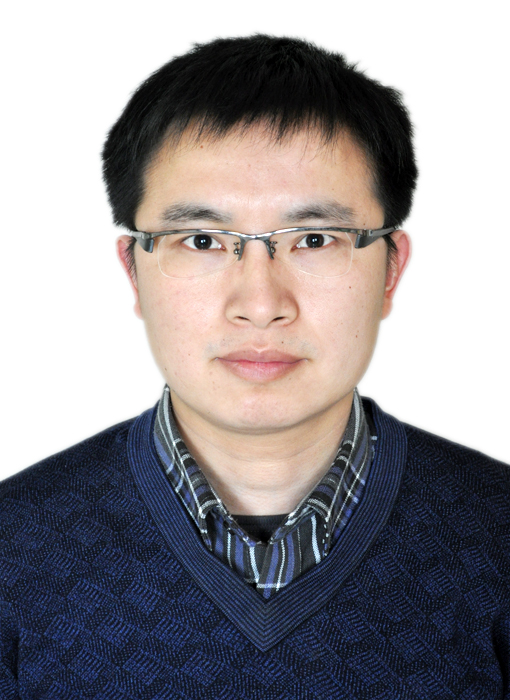}}]{Xuexian Hu} 
is an associate professor in the State Key Laboratory of Mathematical Engineering and Advanced Computing. He received the PhD degree in Information Security from Zhengzhou Information Science and Technology Institute, Zhengzhou, China, in 2010. His current research interests include applied cryptography and big data security.
\end{IEEEbiography}

\begin{IEEEbiography}[{\includegraphics[width=1in,height=1.25in,clip,keepaspectratio]{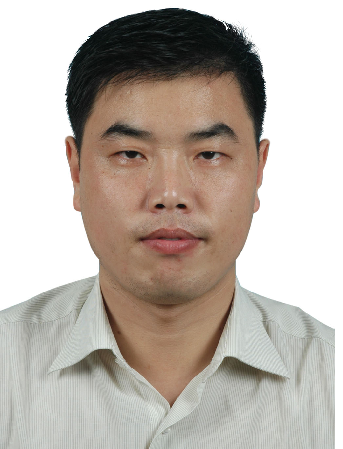}}]{Xiaofeng Chen}
(SM’16) received his B.S. and M.S. on Mathematics from Northwest University, China in 1998 and 2000, respectively. He got his Ph.D degree in Cryptography from Xidian University in 2003. Currently, he works at Xidian University as a professor. His research interests include applied cryptography and cloud computing security. He has published over 200 research papers in refereed international conferences and journals. His work has been cited more than 7000 times at Google Scholar. He is in the Editorial Board of IEEE Transactions on Dependable and Secure Computing, Security and Privacy, and Computing and Informatics (CAI) etc. He has served as the program/general chair or program committee member in over 30 international conferences.
\end{IEEEbiography}

\begin{IEEEbiography}[{\includegraphics[width=1in,height=1.25in,clip,keepaspectratio]{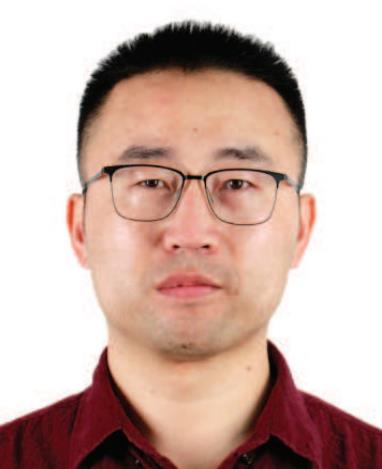}}]{Jianghong Wei} 
received the PhD degree in Information Security from Zhengzhou Information Science and Technology Institute, Zhengzhou, China, in 2016. He is currently a lecturer in the State Key Laboratory of Mathematical Engineering and Advanced Computing, Zhengzhou, China. His research interests include applied cryptography and network security.
\end{IEEEbiography}

\begin{IEEEbiography}[{\includegraphics[width=1in,height=1.25in,clip,keepaspectratio]{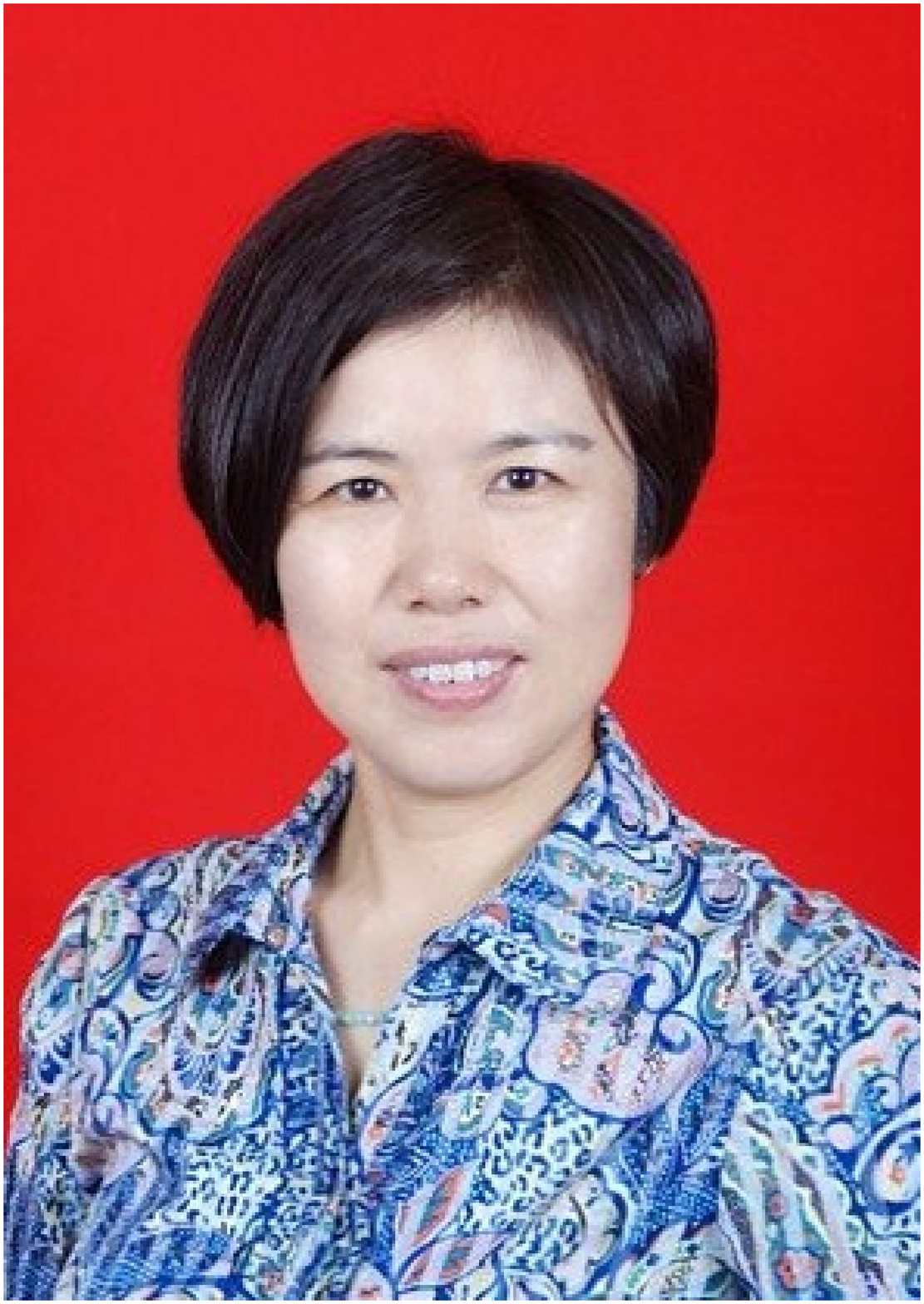}}]{Wenfen Liu}
received the PhD degree in mathematics from Institute of Information Engineering, Zhengzhou, China, in 1998. She is a full professor in Guangxi Key Laboratory of Cryptogpraphy and Information Security, School of Computer Science and Information Security, Guilin University of Electronic Technology, Guilin, China, and serves as the head of probability statistics. Her research interests include probability statistics, network communications, and information security.
\end{IEEEbiography}

\end{document}